\documentclass[sigconf,twocolumn,10pt]{acmart}
\acmSubmissionID{}
\renewcommand\footnotetextcopyrightpermission[1]{}
\usepackage{pifont}
\usepackage{graphicx} 
\usepackage{amsmath}
\usepackage{multirow}
\usepackage{subfigure}
\usepackage{makecell}
\usepackage[ruled,linesnumbered, noend]{algorithm2e}
\usepackage{enumitem}
\usepackage{amsmath}
\usepackage{xcolor}
\usepackage{lipsum}
\usepackage{comment}

\title{MNN-AECS: Energy Optimization for LLM Decoding on Mobile Devices via Adaptive Core Selection} 
\date{April 2025}

\author{Zhengxiang Huang}
\orcid{0009-0000-1536-1616}
\affiliation{%
  \institution{Shanghai Jiao Tong University}
  \city{Shanghai}
  \country{China}
}
\email{huangzhengxiang@sjtu.edu.cn}

\author{Chaoyue Niu}
\authornote{Chaoyue Niu is the corresponding author (rvince@sjtu.edu.cn).}
\orcid{0000-0002-1650-4233}
\affiliation{%
  \institution{Shanghai Jiao Tong University}
  \city{Shanghai}
  \country{China}
}
\email{rvince@sjtu.edu.cn}

\author{Zhaode Wang}
\affiliation{%
  \institution{Alibaba Group}
  \city{Hangzhou}
  \country{China}
}

\author{Jiarui Xue}
\affiliation{%
  \institution{Shanghai Jiao Tong University}
  \city{Shanghai}
  \country{China}
}

\author{Hanming Zhang}
\affiliation{%
  \institution{Shanghai Jiao Tong University}
  \city{Shanghai}
  \country{China}
}

\author{Yugang Wang}
\affiliation{%
  \institution{Shanghai Jiao Tong University}
  \city{Shanghai}
  \country{China}
}

\author{Zewei Xin}
\affiliation{%
  \institution{Shanghai Jiao Tong University}
  \city{Shanghai}
  \country{China}
}

\author{Xiaotang Jiang}
\orcid{0009-0006-9425-7289}
\affiliation{
  \institution{Alibaba Group}
  \city{Hangzhou}
  \state{Zhejiang}
  \country{China}
}

\author{Chengfei Lv}
\orcid{0009-0004-6618-521X}
\affiliation{
  \institution{Alibaba Group}
  \city{Hangzhou}
  \state{Zhejiang}
  \country{China}
}

\author{Fan Wu}
\orcid{0000-0003-0965-9058}
\affiliation{%
  \institution{Shanghai Jiao Tong University}
  \city{Shanghai}
  \country{China}
}

\author{Guihai Chen}
\orcid{0000-0002-6934-1685}
\affiliation{%
  \institution{Shanghai Jiao Tong University}
  \city{Shanghai}
  \country{China}
}

\begin{document}

\begin{abstract}
As the demand for on-device Large Language Model (LLM) inference grows, energy efficiency has become a major concern, especially for battery-limited mobile devices. Our analysis shows that the memory-bound LLM decode phase dominates energy use, and yet most existing works focus on accelerating the prefill phase, neglecting energy concerns. We introduce Adaptive Energy-Centric Core Selection (AECS) and integrate it into MNN to create the energy-efficient version, MNN-AECS, the first engine-level system solution without requiring root access or OS modifications for energy-efficient LLM decoding. MNN-AECS is designed to reduce LLM decoding energy while keeping decode speed within an acceptable slowdown threshold by dynamically selecting low-power CPU cores. MNN-AECS is evaluated across 5 Android and 2 iOS devices on 5 popular LLMs of various sizes. Compared to original MNN, MNN-AECS cuts down energy use by 23\% without slowdown averaged over all 7 devices and 4 datasets. Against other engines, including llama.cpp, executorch, mllm, and MediaPipe, MNN-AECS delivers 39\%–78\% energy saving and 12\%–363\% speedup on average.
\end{abstract}



\maketitle

\section{Introduction}
\label{introduction}

\subsection{Importance of On-Device LLM Inference}
The daily demands for Large Language Model (LLM) service are rapidly growing, with most serving workload being processed on cloud servers currently \cite{DistServe, HCache}. Yet, such on-cloud LLM service incurs high expenses to LLM service providers and raises privacy concerns to end users. 

By contrast, on-device LLM inference as an alternative solution addresses these 2 issues: it significantly relieves workload on cloud servers and reduces serving expenses, and it's also much more privacy-preserving because no queries and generated texts are ever exposed to the servers. These advantages have garnered increasing attention from both LLM teams and mobile phone companies. Qwen2.5 series \cite{Qwen2.5} provide distilled small models (0.5B, 1.5B, 3B) for on-device use cases in MNN \cite{Walle} and GGUF \cite{llama.cpp, GGUF} format. Llama3.2 \cite{Llama3.2} collaborates with executorch \cite{executorch} to launch on-device inference pipeline of its 1B/3B models. Google launches Gemini Nano (2B) with AI Edge SDK \cite{Gemini-Nano}. Apple Intelligence \cite{Apple-Intelligence} adopts device-cloud collaborative serving framework \cite{Apple-Intelligence-Device-Cloud}, where a 3B model handles simple tasks on mobile devices to relieve cloud server workload. 


\begin{table}[!t]
    \centering
    \caption{Energy consumption and CPU utilization of Qwen2.5-1.5B (4-bit quantized), running 20 data entries of multi-turn conversation sampled from ShareGPT dataset, with inference engine MNN.}
    \vspace{-0.5em}
    \resizebox{0.95\columnwidth}{!}{
    \begin{tabular}{cccccc}
    \toprule[1.5pt]
        \multirow{2}{*}{\bf device} & {\bf total} & \multicolumn{2}{c}{\bf consumption} & \multirow{2}{*}{\bf CPU use} & \multirow{2}{*}{\bf power} \\
        \cmidrule(r){3-4}
        & {\bf battery} & {\bf battery} & {\bf energy} & & \\
        \midrule
        Xiaomi 15 Pro & 6100 mAh & 386 mAh (6\%) &  6031 J & 395\% & 9.9 W \\
        Mate 40 Pro & 4400 mAh & 717 mAh (16\%) & 10438 J & 396\% & 8.7 W  \\
        iPhone 12 & 2815 mAh &  708 mAh (25\%) &  10379 J & 252\% & 7.9 W \\
    \bottomrule
    \end{tabular}
    }
    \label{tab:energy-challenge}
    \vspace{-0.5em}
\end{table}


\begin{figure*}[!t]
    \centering
    \subfigure[The 8 circles in the thread pool box represent 8 cores, and the yellow ones are selected. Bigger circles represent bigger cores in big-little core CPU architecture. MNN-AECS searches and selects less and smaller cores than original MNN in decode phase to reduce CPU utilization and frequency, resulting in lower energy.]{\includegraphics[width=0.7\linewidth]{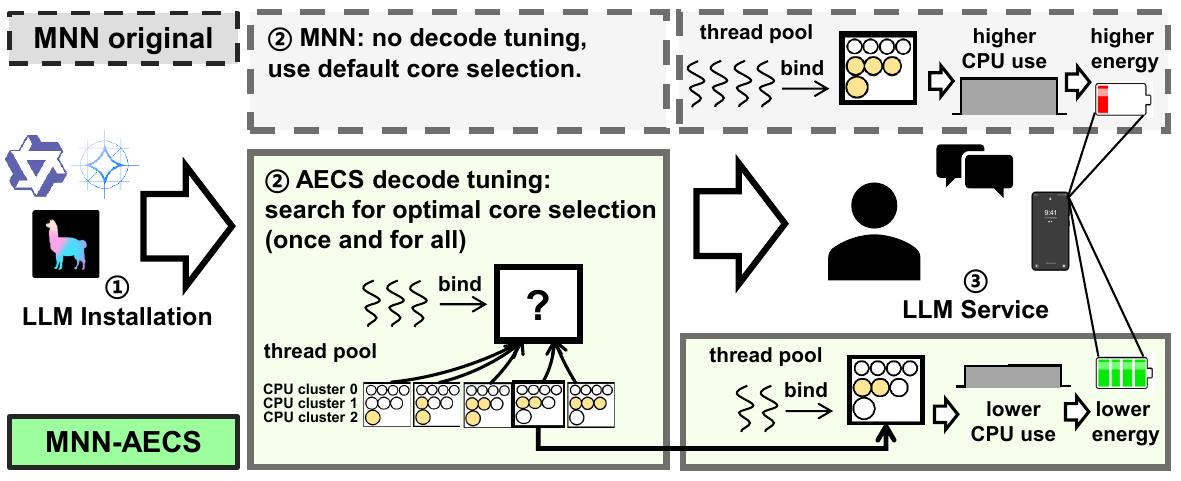}\label{fig:workflow}}
    \hspace{0.8em}
    \subfigure[Energy and speed are presented in ratio relative to MNN-AECS.]{\includegraphics[width=0.23\linewidth]{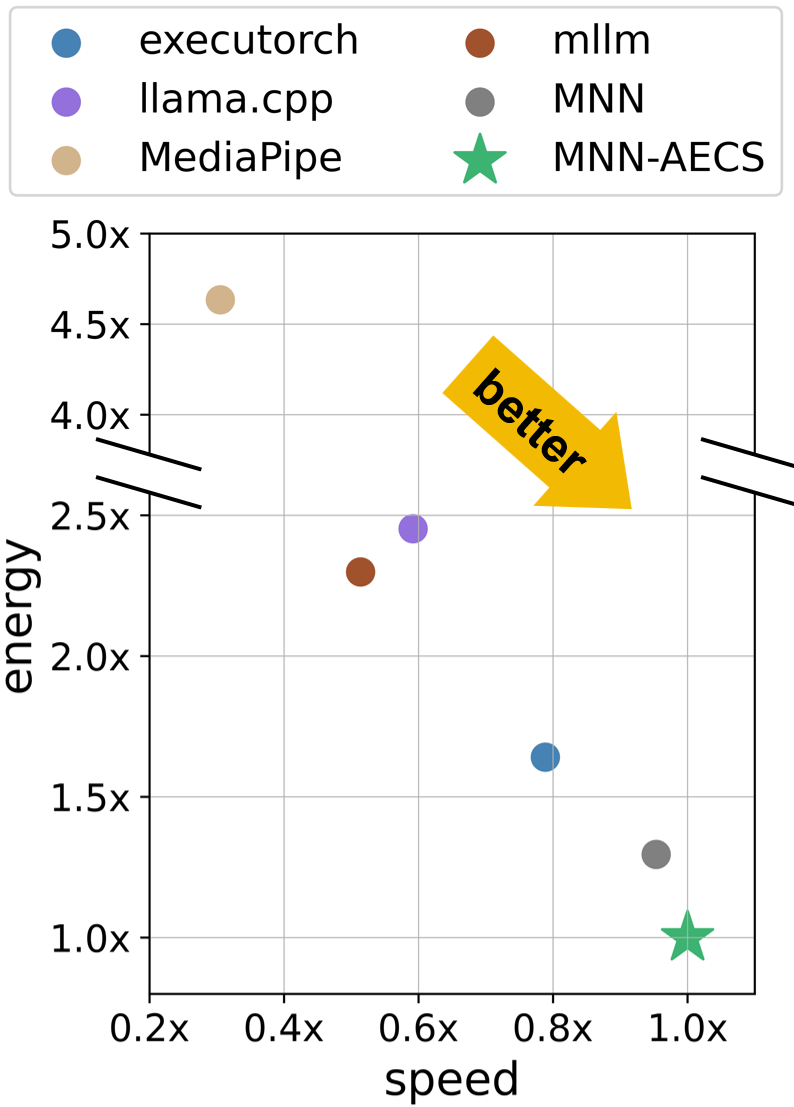}\label{fig:performance}}
    
    \vspace{-1em}
    \caption{(a) MNN-AECS workflow. (b) MNN-AECS geometric mean performance over 5 baseline engines across 7 devices and 4 datasets. MNN-AECS (green star) consumes the lowest energy and is also the fastest.}
    \label{fig:workflow-performance}
    \Description{See caption}
    \vspace{-1em}
\end{figure*}

\subsection{On-Device LLM Inference Energy Issue}
\label{energy-challenge}
Energy is one of the major issues for on-device LLM inference. Heavy workload of LLM inference leads to high hardware utilization and power and prolongs execution time, resulting in intolerably high energy consumption (Table \ref{tab:energy-challenge}), draining mobile device battery and raising CPU temperature.

\textbf{Heavy workload of LLM inference.} \label{heavy-workload} Compared to traditional smaller discriminative Deep Learning (DL) models, LLMs induces much higher energy consumption due to their heavier workload caused by large model sizes and auto-regressive decoding characteristic. 1) Model sizes of mobile LLMs typically scale up to 3B \cite{Qwen2.5, Llama3.2, Gemma2} currently, leading to 3-10 GFLOPS/token computation and up to 1-3 GB/token DRAM memory visits even after 4/8-bit quantization, resulting in heavy per-token workload both on processors and memory bus. Processor utilization is boosted to nearly maximum, raising the device power to as high as 8-10 W, as is shown in Table \ref{tab:energy-challenge}. 2) For each auto-regressive decoding loop, LLM inputs the last previously generated token and outputs a new token one at a time until reaching the end of sentence. Usually hundreds of tokens are generated in each round of LLM decoding, thus  prolonging execution time to minute-level, cumulating the heavy per-token energy consumption to an unprecedentedly high total amount.


\textbf{Limited battery capacity} of mobile devices is an important hardware constraint given the high energy consumption of LLM inference. 
Unlike cloud servers provisioned with continuous power supply, mobile phones, 
for unplugged and ubiquitous use, are equipped with limited incorporated battery. Table \ref{tab:energy-challenge} shows the battery capacity of 3 common mobile devices: Xiaomi 15 Pro (high-end), Huawei Mate 40 Pro and iPhone 12 (middle-end), ranging from 2815 to 6100 mAh.


\textbf{Consequences of high energy consumption.} 1) The battery runs out too fast, which is the direct consequence of high energy consumption. For example, in Table \ref{tab:energy-challenge}, 20 conversations sampled from ShareGPT \cite{ShareGPT} consume up to $6\%\sim 25\%$ battery in less than 15 minutes. Hence, users may suffer from the inconvenience of frequent charging. 2) High temperature can be a malicious side-effect of high cumulative energy consumption. CPU temperature can climb up to $80^\circ C$ during LLM inference. High temperature may degrade user experience due to CPU throttling \cite{MobiRL} and gradually damage the hardware \cite{LowPower, arm-power}.

\subsection{Our Work}
In this work, we target at the energy challenge of on-device LLM inference. Our energy analysis of the two LLM phases, prefill and decode, reveals that decode phase is the primary contributor to the total LLM inference energy consumption on common on-device text generation tasks.
Thus, we focus our design on optimizing LLM decode energy.

We introduce a novel method: Adaptive Energy-centric Core Selection (AECS), to optimize the LLM decode energy with speed guarantee (no more than $\epsilon$ decode slowdown). We also integrate it into on-device LLM  engine MNN \cite{Walle} and propose our energy-efficient version of it: MNN-AECS. 

Figure \ref{fig:workflow} shows the workflow of MNN-AECS. Between installation and  LLM service, we add a once-and-for-all AECS decode tuning, which searches for the optimal core selection that reduces decode energy with negligible speed loss by reducing CPU utilization, where a CPU core selection is a core binding plan on Android and a thread number on iOS. AECS adopts a heuristic search rather than a naive exhaustive one to guarantee the optimality of search results and also reduce search time by up to 90\%. After tuning, the searched optimal core selection is input to the thread pool  as decode configuration for all future services. MNN is modified so that prefill and decode can adopt different core selections, eliminating interference between prefill and decode. 




To position our work, \textbf{MNN-AECS is a pure engine-level system solution} implemented outside OS without root access and also orthogonal to any algorithmic designs.

To evaluate MNN-AECS, we develop a cross-platform LLM energy evaluation testbed to measure energy and speed of on-device LLM engines on both Android and iOS devices. The energy profiling module of the testbed provides precise energy data probed from OS interfaces and passed to  MNN-AECS and evaluation pipeline though JNI on Android and tunneld for iOS. MNN-AECS is compared against original MNN, llama.cpp, executorch, mllm, and MediaPipe across 5 Android and 2 iOS devices, using 5 LLMs: Qwen2.5-1.5B/3B, Llama3.2-1B/3B, and Gemma2-2B. 
Results in Figure \ref{fig:performance} summarize MNN-AECS normalized performance across all devices and datasets, reducing energy by 23\% over its base engine MNN without slowdown, and by 39\%–78\% compared to all other engines.

In summary, the contributions of this paper include: 
\begin{itemize}
    \item We analyze the energy challenge of on-device LLM inference and identify that LLM decoding dominates energy consumption in Section \ref{background}.
    \item We propose MNN-AECS, a novel energy-efficient version of MNN that optimizes LLM decoding energy while preserving speed in Sections \ref{design} and \ref{implementation}.
    \item We develop a cross-platform LLM energy evaluation testbed, and evaluate MNN-AECS against state-of-the-art on-device LLM inference engines in Section \ref{evaluation}.
    \item We explore potential future extensions to mobile XPUs in Section \ref{discussion}.
\end{itemize} 



\section{Background}
\label{background}

\subsection{On-Device LLM Inference}
LLM model architecture consists of multiple sequential transformer blocks, each containing an attention module \cite{attention} and an FFN (Feed Forward Network) \cite{attention} or Gated-MLP \cite{Llama3.2, Qwen2.5, Llama3}. LLM inference pipeline can be divided into 2 phases: prefill phase and decode phase, showing different operational intensity.

\textbf{Compute-bound prefill phase.} Prefill phase takes a user prompt as input, with length $p$ after tokenization. Denote the hidden dimension of LLM as $d$, and then all the intermediate activations $H$ are of the shape $p \times d$, being matrices. For attention modules, these matrices $H$ are input and linearly affined to 
\begin{equation}
\begin{aligned} 
    Q,K,V=H\cdot W^{Q,K,V},\ H\in \mathbb{R}^{p\times d}.
\end{aligned}
\label{prefill-QKV}
\end{equation}
Then, self-attention inputs Q, K, V, and computes as
\begin{equation}
\begin{aligned}
    \mathbf{attention}(H)=\mathbf{softmax}(\mathbf{mask}(\frac{Q\cdot K^T}{\sqrt{d}}))\cdot V.
\end{aligned}
\label{prefill-attn}
\end{equation}
Subsequently, FFN/GatedMLP are formulated as
\begin{equation}
    \begin{aligned}
        &\mathbf{FFN}(H)=F_{act}(H\cdot W_{up} + b_{up})W_{down}+b_{down},\\
        &\mathbf{GatedMLP}(H)=(F_{act}(H\cdot W_{gate})*(H\cdot W_{up}))W_{down}, 
    \end{aligned}
    \label{prefill-mlp}
\end{equation}
where the input matrix $H$ is multiplied by weight matrices.

Both prefill attention and FFN/GatedMLP are dominated by Matrix-Matrix Multiplication (GEMM). Hence, LLM prefill phase is compute-bound.

\textbf{Memory-bound decode phase.}  During decoding, $K,V$ tensors of previous tokens are cached in a structure called KV cache for reuse, not needed to compute again \cite{vllm}. Therefore, only 1 token needs to be processed during each decoding, causing the shape of intermediate activations $h$ to be $1\times d$, a vector. For each attention module, vectors $q,k,v$ are computed from $h$, and $k,v$ are concatenated to previous KV matrices $K_C$ and $V_C$, formulated as
\begin{equation}
\begin{aligned}
    &q,k,v=h\cdot W^{Q,K,V},\ h\in \mathbb{R}^{1\times d},\\
    &K_c:=\begin{bmatrix}
 K_c\\
 k
 \end{bmatrix}, V_c:=\begin{bmatrix}
 V_c\\
 v
 \end{bmatrix}.\\
\end{aligned}
\label{KV-cache}
\end{equation}
Then, Self-attention is performed between vector $q$ and matrices $K_C,V_C$ as follows:
\begin{equation}
\begin{aligned}
    &\mathbf{attention}(h)=\mathbf{softmax}(\frac{q\cdot K_c^T}{\sqrt{d}})\cdot V_c. \\
\end{aligned}
\label{decode-attn}
\end{equation}
After that, the intermediate state vector $h$ is input to: 
\begin{equation}
    \begin{aligned}
        &\mathbf{FFN}(h)=F_{act}(h\cdot W_{up} + b_{up})W_{down}+b_{down},\\
        &\mathbf{GatedMLP}(h)=(F_{act}(h\cdot W_{gate})*(h\cdot W_{up}))W_{down}, 
    \end{aligned}
\label{decode-mlp}
\end{equation}
where vector $h$ is multiplied by weight matrices.

Compared with prefill operations in Equations \ref{prefill-QKV}, \ref{prefill-attn},  \ref{prefill-mlp}, matrix $H$ is replaced with vector $h$,  and $Q$ is also replaced with vector $q$, so that both decode attention and FFN/Gated-MLP are dominated by Matrix-Vector Multiplication (GEMV), and thus memory-bound. 

\begin{table*}[t]
    \centering
    \caption[Mobile devices used in our experiments and their hardware and OS specification.]{Mobile devices used in our experiments and their hardware and OS specification. \footnotemark}
    \vspace{-0.5em}
\resizebox{0.86\linewidth}{!}{
    \begin{tabular}{cc|cc|cccc|cc}
    \toprule[1.5pt]
         & {\bf device} & {\bf DRAM} & {\bf battery} & {\bf SoC} & {\bf CPU} & {\bf GPU} & {\bf NPU} & {\bf OS} & {\bf freq scheduler}\\
         \midrule
         \multirow{15}{*}{\bf Android} & Huawei & \multirow{3}{*}{8GB} & \multirow{3}{*}{4400 mAh} & Hisilicon & 1*A77(3.13GHz) & Mali-G78 & Da Vinci & HarmonyOS & \multirow{3}{*}{schedutil} \\
         & Mate 40 Pro & & & Kirin 9000 & +3*A77(2.54GHz) & (24 cores) & (2+1) & 4.2.0 &  \\
         & & & & & +4*A55(2.05GHz) &  &  &  &  \\
         \cmidrule(r){6-6}
         & \multirow{3}{*}{Honor V30 Pro} & \multirow{3}{*}{8GB} & \multirow{3}{*}{4100 mAh} & Hisilicon  & 2*A76(2.86GHz)
 & Mali-G76 & Da Vinci & HarmonyOS & \multirow{3}{*}{schedutil} \\
         & & & & Kirin 990 & +2*A76(2.36GHz) & (16 cores) & (2+1) & 4.2.0 &  \\
         & & & & & +4*A55(1.95GHz) &  &  &  &  \\
         \cmidrule(r){6-6}
         & Samsung & \multirow{3}{*}{8GB} & \multirow{3}{*}{5000 mAh} & \multirow{3}{*}{Exynos 1580} & 1*A720(2.9GHz) & Xclipse 540  & \multirow{3}{*}{1 core} & \multirow{3}{*}{One UI 7.0} & \multirow{3}{*}{schedutil} \\
          & Galaxy A56 & & &  & +3*A720(2.6GHz) & (2 cores, 256 shaders) &  &  &  \\
         & & & & & +4*A520(1.95GHz) &  &  &  &  \\
         \cmidrule(r){6-6}
         & \multirow{4}{*}{Meizu 21} & \multirow{4}{*}{12GB} & \multirow{4}{*}{4800 mAh} &   & 1*X4(3.3GHz)   & \multirow{4}{*}{Adreno 750} & \multirow{4}{*}{Hexagon} &  & \multirow{4}{*}{walt} \\
         & & & & Snapdragon & +3*A720(3.15GHz) & & & FlymeOS &  \\
         & & & & 8 Gen 3 & +2*A720(2.96GHz) &  &  & 11.2.0 &  \\
         & & & & & +2*A520(2.27 GHz) &  &  &  &  \\
         \cmidrule(r){6-6}
         & \multirow{2}{*}{Xiaomi 15 Pro} & \multirow{2}{*}{16GB} & \multirow{2}{*}{6100 mAh} & Snapdragon  &  2*Oryon(4.32GHz)  & \multirow{2}{*}{Adreno 830} & \multirow{2}{*}{Hexagon} & HyperOS & \multirow{2}{*}{walt} \\
         & & & & 8 Elite & +6*Oryon(3.53GHz) &  &  & 2.0.23 &  \\
         \midrule
         \multirow{2}{*}{\bf Apple} & iPhone 12 & 4GB & 2815 mAh & Apple A14 & 2+4 & 4 cores & 16 cores & iOS 18.3.2 &  \\
         & iPhone 15 & 8GB & 3349 mAh & Apple A16  & 2+4 & 5 cores &  & iOS 18.4.1 &  \\
    \bottomrule[1.5pt]
    \end{tabular}
    }
    \vspace{-0.8em}
    \label{tab:devices}
\end{table*}

\footnotetext{All the Android CPU start with A or X stands for ARM Cortex-A or ARM Cortex-X. Some of the iPhone information is inaccessible and left blank.} 

Moreover, given that normally on-device LLM serves only 1 user and thus processes 1 query at a time (i.e., batch size = 1), decoding can't be batched \cite{continuous-batching, Orca} to remove its memory-boundedness. Therefore, we conclude that on-device LLM decoding is memory-bound.

\subsection{Decoding Dominates Energy}
\label{decode-dominance}
\begin{figure}[!h]    
    \centering
    \vspace{-0.6em}
    \includegraphics[width=0.95\linewidth]{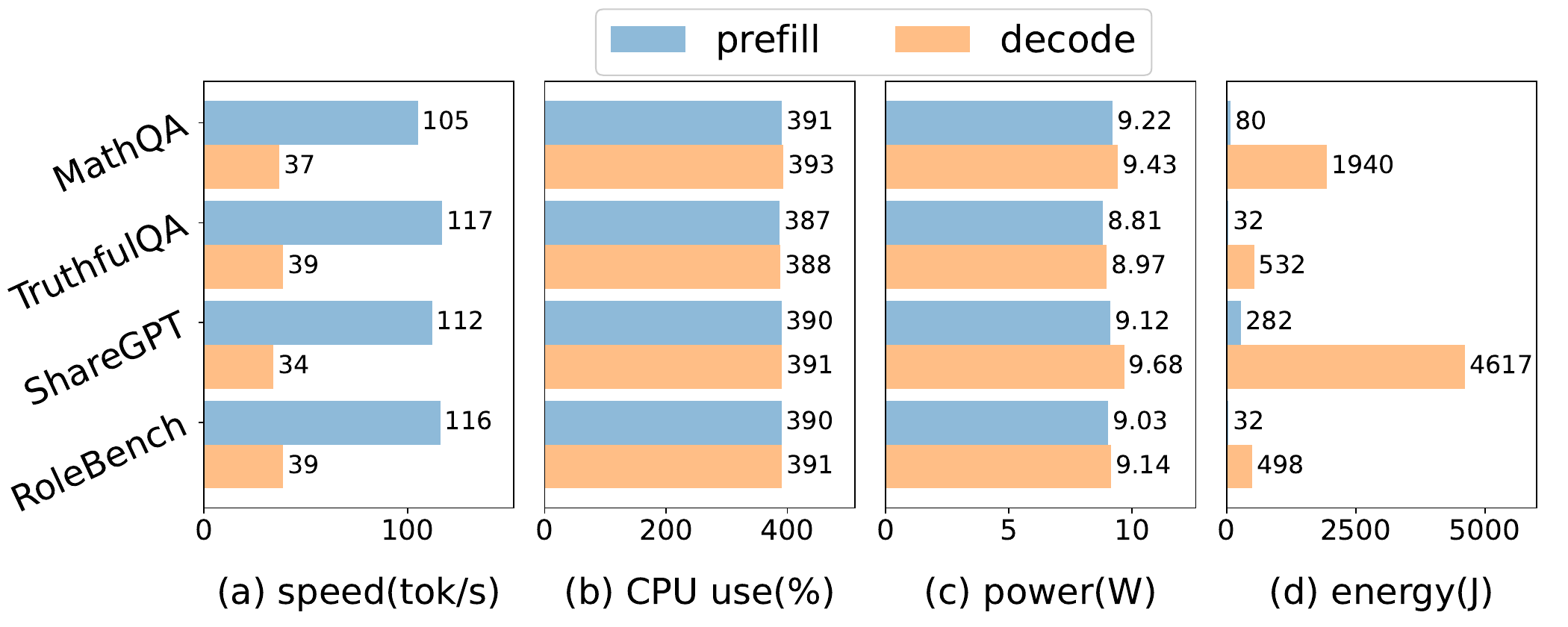}
    \vspace{-0.8em}
    \caption{Comparison of prefill and decode speed, CPU use, power, and energy of Qwen2.5-1.5B across 4 datasets on Xiaomi 15 Pro.}
    \Description{See text}
    \label{fig:decode-dom}
    \vspace{-0.8em}
\end{figure}

Energy consumption is the product of time and power.
Among the 2 LLM phases, decode phase has a longer execution time and similar power compared to prefill, dominating the overall LLM inference energy use. 
Figure \ref{fig:decode-dom}(d) shows that decode energy is 16$\times$ to 26$\times$ more than prefill.


\textbf{Decode phase takes much longer time,} in that decode speed is slower and decode length is longer. 1) Decode speed is  3$\times$ slower than prefill on Xiaomi 15 Pro CPU as an example in Figure \ref{fig:decode-dom}(a). Due to memory-boundedness, all XPU cores share and wait for the same underlying memory bus on mobile devices under Unified Memory Architecture (UMA), making decoding hard to speed up at system level. By contrast, compute-bound prefill can be accelerated well by workload balance \cite{Asymo}, layout transformation \cite{SmartMem}, and XPU \cite{mllm, PowerInfer-2}.

\begin{figure}[!h]
    \centering
    \vspace{-0.5em}
    \subfigure[ShareGPT]{\includegraphics[width=0.41\columnwidth]{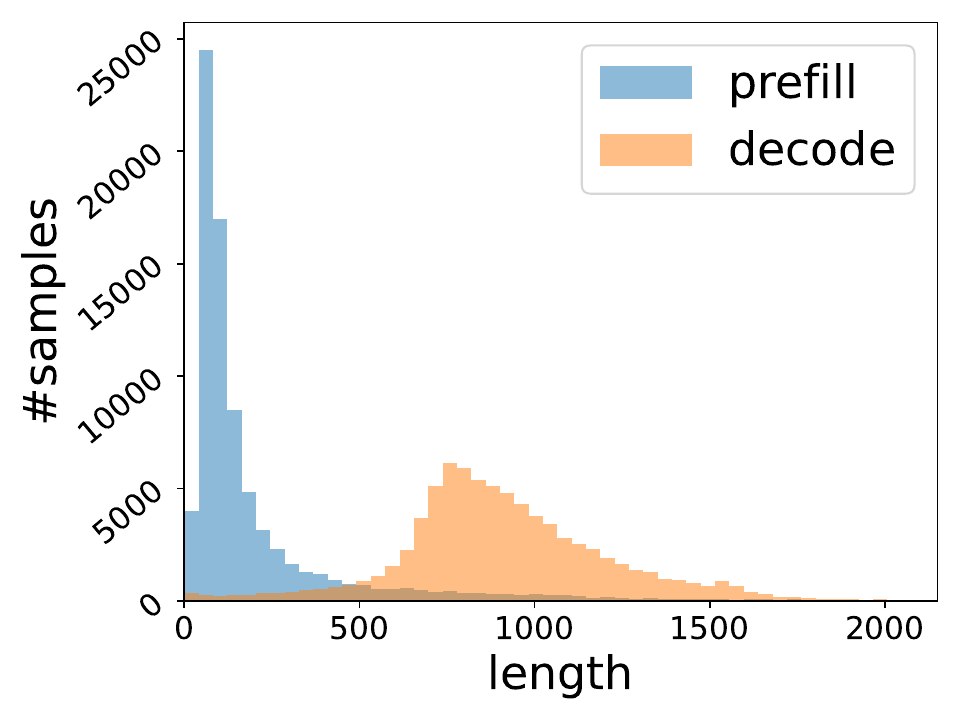}}
    \hspace{1em}
    \subfigure[RoleBench]{\includegraphics[width=0.41\columnwidth]{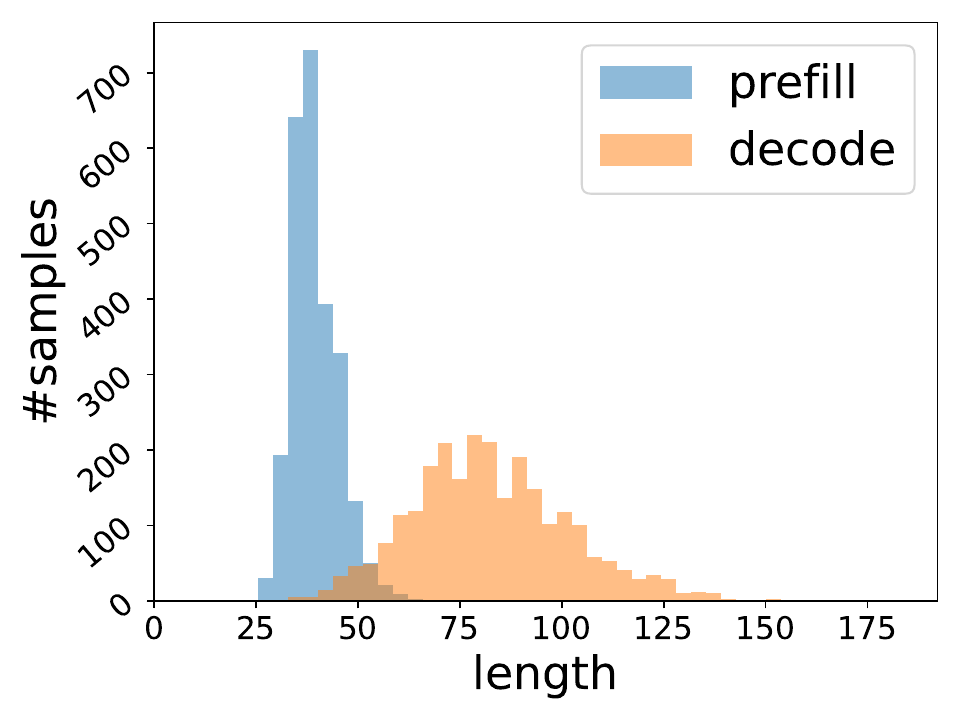}}
    
    \vspace{-0.8em}
    \caption{Prefill length and decode length distributional characteristic. (decode length $\approx$ 3.5$\times$ prefill length)}
    \Description{See Caption}
    \vspace{-0.7em}
    \label{fig:distribution}
\end{figure}

\noindent 2) Decode length is 3.5$\times$ longer than prefill length on 
conversational text generation such as ShareGPT \cite{ShareGPT} and RoleBench \cite{RoleBench}, shown in Figure \ref{fig:distribution}. This is because user queries (prefill) are usually shorter than generated responses (decode).

\textbf{Decode phase power is comparable to prefill.}  Figures \ref{fig:decode-dom}(b) and \ref{fig:decode-dom}(c) show that CPU utilization is similarly high during both prefill and decode phases, and so is the power, as power is positively correlated with CPU utilization. 
This phenomenon is due to existing engines' strategies of utilizing the same number of CPU cores in both phases.


Thus, we \textbf{focus on decode phase energy optimization}, especially reducing decode power, because decode time can hardly speed up due to memory-boundedness. Decode power can be reduced by decreasing CPU utilization during decode via leveraging mobile hardware and OS characteristics.

\subsection{Mobile Hardware Characteristics}
Multi-core XPU is a major hardware characteristic of smartphones from mainstream brands \cite{iPhone, Samsung, Xiaomi, Huawei, Meizu}, which motivates our AECS design: adaptive selection across CPU multi-cores, introduced in Section \ref{design}. We also clarify the rationale of utilizing CPU for LLM decoding in our design.

\textbf{Multi-core architecture of mobile XPU.} Table \ref{tab:devices} presents the devices for evaluation in this work. All of them feature multi-core XPU. For CPU, they consist of several heterogeneous multi-core clusters, e.g., ARM big.LITTLE core CPU architecture \cite{big.Little}. Such heterogeneity aims at assigning only less power-consuming clusters for small bursty workloads to save energy. GPUs such as Adreno and Mali also consist of a series of cores, each containing hundreds of shaders \cite{Snapdragon, Mali}. NPUs are specialized in neural network operations, especially Matrix Multiplication. Huawei Da Vinci series also feature heterogeneous cores for different kinds of neural network operations.


\begin{table}[!h]
    \centering
    \vspace{-0.2em}
    \caption{Core Configurability of mobile XPUs.}
    \vspace{-0.6em}
    \resizebox{0.7\columnwidth}{!}{
    \begin{tabular}{cccc}
    \toprule[1.5pt]
         & \textbf{CPU} & \textbf{GPU} & \textbf{NPU} \\
        \hline
        core frequency & \ding{53} & \ding{53} & \ding{53} \\
        core number & \ding{52} & \ding{53} & \ding{53} \\
        \multirow{2}{*}{core affinity} & Android: \ding{52}  & \multirow{2}{*}{\ding{53}} & \multirow{2}{*}{\ding{53}} \\
        & iOS: \ding{53} & &  \\
    \bottomrule[1.5pt]
    \end{tabular}
    }
    \vspace{-0.5em}
    \label{tab:programmability}
\end{table}

\textbf{Rationale of using CPU for LLM decoding.} 
1) CPU is the most configurable backend outside OS without root access, shown in Table \ref{tab:programmability}. It offers us the largest design space and opportunity. Thread number and even affinity (on Android) can be designated by LLM engine, which is impossible on GPU/NPU. These interfaces can be utilized to search for energy-efficient core selection. 2) CPU decoding speed is comparable to mobile GPU/NPU recently due to memory-boundedness, so that most existing engines majorly utilize CPU decoding \cite{llama.cpp, Walle, mllm, PowerInfer-2}. CPU's configurability and fast decoding speed motivates us to design energy-efficient CPU decoding.

\subsection{Mobile OS Characteristics}
\label{OS}
Mobile OS is capable of manipulating hardware frequency based on process elapsed time for general-purpose energy saving, but not suitable for LLM energy optimization. Outside OS without root access, only CPU affinity or thread number setting are possible, yet still being our design opportunity.

\textbf{Inside OS exists general-purpose energy saving via frequency scheduling, but incapable of reducing LLM decoding energy.} By controlling the operating frequency of mobile XPUs, OS DVFS (Dynamic Voltage Frequency Scaling), specifically CPUFreq \cite{CPUFreq} and devfreq \cite{devfreq} subsystems, adopt governors such as schedutil \cite{schedutil} and walt \cite{walt} to schedule frequency and save energy when system estimated workload is low \cite{MobiRL}. Workload is estimated based on tasks elapsed time. However, such workload-based scheduling is incapable of reducing LLM decode phase energy, because it can't distinguish decode from prefill and leverage the memory-boundedness of LLM decoding to scale down frequency and consequently save energy without speed loss.

\textbf{Outside OS exist LLM decoding energy saving opportunities through direct frequency setting with root access and through core selection system calls without root access.} 
With root access, it is possible to directly set the core frequency to reduce decode energy. However, obtaining root access outside the OS is impractical for commercial use. Without root access, we leverage core selection interfaces instead, including CPU affinity and thread number setting. 
On Android, CPU affinity (binding threads to designated cores) is available. On iOS, though affinity is unavailable, thread number can be controlled. With proper core selection via core binding and thread number setting, we can leave cores idle during memory-bound decoding, which can lead to significant energy reduction without intolerable speed loss, as detailed in Section \ref{rationale}. This core selection approach represents a key design opportunity for our work.

\subsection{Low-Power Consideration in Existing On-Device LLM Inference Engines}
\label{background-engine}
Given the practical challenge of reducing energy consumption of LLM inference, on-device LLM inference engine as a middle-ware is motivated to leverage  hardware and OS characteristics to encounter it. However, current on-device LLM frameworks lack sufficient focus on low-power optimization. For example, MNN \cite{Walle} offers only a \textit{Power\_Low} mode that forces single-thread execution, which cannot meet the speed demands of LLM inference. Other engines, such as llama.cpp\cite{llama.cpp}, executorch \cite{executorch}, MediaPipe \cite{MediaPipe}, mllm \cite{mllm}, and PowerInfer-2 \cite{PowerInfer-2}, focus on accelerating the prefill phase to reduce prefill energy but fail to address the more energy-consuming decode phase. This is because their optimizations do not apply to decoding, which is memory-bound. Therefore, an engine-level design is in need to enable energy-efficient LLM decoding.



\section{Design}
\label{design}

\subsection{Optimization Objective}


Our major objective is to reduce LLM decoding energy, the bottleneck. Meanwhile, decode speed also matters for user experience. Hence, \textbf{we optimize energy under speed constraint}. We formulate the optimization objective as follows:
\begin{equation}
\begin{aligned}
    \min_{I\in S}&\ E(I) = P(I)\cdot t(I),\ t(I)\propto \frac{1}{speed(I)}\\
    &\quad s.t.\frac{speed(I)}{\max\limits_J speed(J)} \ge 1-\epsilon.
\end{aligned}
    \label{optimization}
\end{equation}
Core selection $I$ serves as the decision variable and is chosen from the search space $S$. The objective $E(I)$ is the empirical energy consumption associated with $I$. $E(I)$ is computed as the product of power $P(I)$ and time $t(I)$. The constraint term ensures that the decode speed $speed(I)$ of any feasible $I$ shall be at most $\epsilon$ slower than the fastest decode speed $\max_J speed(J)$. We empirically select $\epsilon=8\%$, which is a slowdown not noticeable to smartphone users. 

In summary, the optimal core selection $I^*$ of Equation \ref{optimization} is the most energy efficient core selection with an  $\epsilon$-suboptimal decode speed guarantee. 

\begin{figure}[!t]
    \centering
    \subfigure[\textbf{llama.cpp}:  all cores are selected.]{\includegraphics[width=0.75\columnwidth]{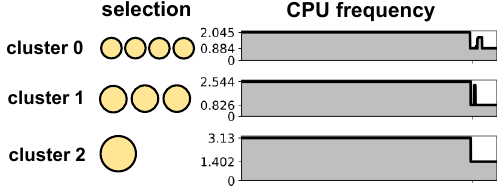}}

    \vspace{-0.4em}
    \hspace{0.2em}\subfigure[\textbf{MNN}: cluster 2 and 1 are selected.]{\includegraphics[width=0.74\columnwidth]{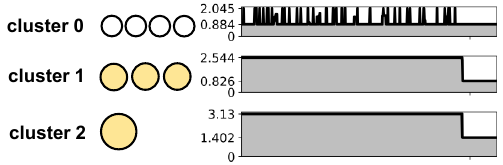}}

    \vspace{-0.4em}
    \hspace{0.6em}\subfigure[\textbf{MNN-AECS}: 2 cores from cluster 1 is selected.]{\includegraphics[width=0.74\columnwidth]{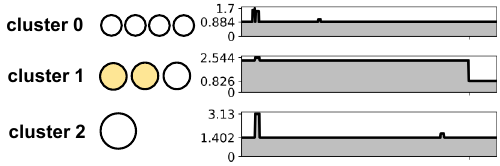}}
    \vspace{-0.5em}
    \caption{CPU frequency curve of 3 core selections of LLM decoding on Huawei Mate 40 Pro.}
    \Description{See Caption}
    \label{fig:android-rationale}
    \vspace{-1em}
\end{figure}

\subsection{Search Space Design}
\label{rationale}

The search space $S$ in Equation \ref{optimization} is all CPU core binding plans on Android and all possible thread numbers on iOS. 
For example, Mate 40 Pro has 3 clusters, from big (cluster 0) to small (cluster 2), containing 1, 3, and 4 cores respectively. The design space of Mate 40 Pro spans all possible combinations of these 8 cores. In contrast, iPhone 12 has 6 cores, and its design space is defined by the thread number ranging from 1 to 6.
We first conduct preliminary experiments to verify such search space design, demonstrating that the optimized core selection can indeed reduce CPU utilization and frequency, and consequently the power in decode phase.

\textbf{Rationale of Android search space design.} 
Figure \ref{fig:android-rationale} shows the preliminary experiments on Mate 40 Pro. llama.cpp selects all 8 cores available including efficient cores, resulting in peaked frequencies across all clusters. MNN selects all 4 cores in cluster 2 and 1, and the frequency of these 2 selected clusters are boosted to peak. By contrast, the optimal solution (discovered by MNN-AECS) only selects 2 cores form cluster 1, successfully reducing the average frequency of idle clusters by 50\%. 
The numerical results presented in Table \ref{tab:android-rationale} indicate that the frequency reduction contributes to 29\% and 65\% energy reduction over MNN and llama.cpp respectively. Besides, our optimal solution only slows down 5\% compared to MNN, because of memory-boundedness. 

The results indicate that a proper CPU binding can leave CPU cores or even clusters idle, reducing frequency and power consumption of them. 

\begin{table}[!h]
    \centering
    \vspace{-0.5em}
    \caption{Average CPU frequency, speed, power, and energy of 3 engines on Mate 40 Pro, Qwen2.5-1.5B.}
    \vspace{-0.5em}
    \resizebox{\columnwidth}{!}{
    \begin{tabular}{ccccccc}
    \toprule[1.5pt]
        \multirow{2}{*}{\textbf{engine}}  & \multicolumn{3}{c}{\textbf{CPU frequency (GHz)}} & \textbf{speed} & \textbf{power} & \textbf{energy} \\
        \cmidrule{2-4}
        & \textbf{cluster 2}& \textbf{cluster 1}& \textbf{cluster 0} & \textbf{(tok/s)} & \textbf{(W)} & \textbf{(mJ/tok)} \\
    \midrule
        llama.cpp & 3.08 & 2.50 & 2.01 & 10.2 & 8.8 & 860 \\
        MNN & 3.05 & 2.47 & 1.04 & \underline{\textbf{21.7}} & 8.7 & 403 \\
        MNN-AECS & \underline{\textbf{1.61}} & \underline{\textbf{2.33}} & \underline{\textbf{0.96}} & 20.6 & \underline{\textbf{6.2}} & \underline{\textbf{300}}\\
    \bottomrule[1.5pt]
    \end{tabular}
    }
    \label{tab:android-rationale}
    \vspace{-0.5em}
\end{table}

\textbf{Rationale of iOS search space design.} Table \ref{tab:iOS-rationale} shows the preliminary experiments on iPhone 12. 
Our optimal solution, selecting only 1 thread, is 42\% and 48\% more energy-saving and 14\% and 100\% faster than MNN and llama.cpp, by saving over 2$\times$ more CPU resources. 
Hence, a properly set thread number can also leaves cores or even clusters idle.

\begin{table}[!h]
    \centering
    \vspace{-0.5em}
    \caption{Average CPU frequency, speed, power, and energy of 3 engines on iPhone 12, Qwen2.5-1.5B.}
    \vspace{-0.5em}
    \resizebox{0.7\columnwidth}{!}{
    \begin{tabular}{ccccc}
    \toprule[1.5pt]
        \multirow{2}{*}{\textbf{engine}} & \multirow{2}{*}{\textbf{thread}} & \multirow{2}{*}{\textbf{CPU use}} & \textbf{speed} & \textbf{relative} \\
        & &  & \textbf{(tok/s)}  & \textbf{power} \\
    \midrule
        llama.cpp & 2 & 197\% & 15.3 & 955  \\
        MNN & 4 & 230\% & 27.6 & 871 \\
        MNN-AECS & \underline{\textbf{1}} & \underline{\textbf{97\%}} & \underline{\textbf{31.5}} & \underline{\textbf{506}} \\
    \bottomrule[1.5pt]
    \end{tabular}
    }
    \vspace{-0.5em}
    \label{tab:iOS-rationale}
\end{table}

These preliminary experiments demonstrate how the optimal core selection saves considerable energy without losing much speed or even faster than existing works, as long as such solution can be found by our searching algorithm.


\subsection{Adaptive Energy-Centric Core Selection (AECS)}
We present our algorithm AECS: search for a core selection that optimizes Equation \ref{optimization} based on CPU information and the once-and-for-all on-device searching.

Algorithm \ref{alg:AECS} shows the workflow of AECS. It has 2 search stages. \textbf{Stage 1} searches for the fastest core selection $\tilde{I}$ so that following searches can compare with speed of it $speed(\tilde{I})$ check the speed constraint. It also serves as the initializer in stage 2 search. \textbf{Stage 2} introduces a heuristic power estimation function $h(I)$ to guide the search. Based on stage 1 result and the heuristic function, stage 2 generates a heuristically pruned candidate tree $S_h(\tilde{I})$ as the candidate set. Then each candidate core selection on the tree is tested and profiled. Finally, the best candidate core selection $I^*$ minimizing the heuristically modified energy function $E_h(\cdot)$ and satisfying the speed constraint is returned.

\textbf{Stage 1: searching for the fastest core selection.} We first search for the core selection $\tilde{I}$ that achieves the fastest decoding, so that we can check and verify the speed constraint in the following stage 2 energy optimization search. 

\begin{algorithm}[!t]
  \caption{AECS} 
  \label{alg:AECS}
  \small
  \SetKwInOut{KwInput}{Input}
  \SetKwInOut{KwOutput}{Output}
  
  \KwInput{CPU information.}
  \KwOutput{The optimized core selection $I^*$.}

    Read and analyze CPU info.
    
    \textbf{// Stage 1:} 
    
    Search for the fastest core selection $\tilde{I}$.

    \textbf{// Stage 2:}
    
    Generate candidate set $T_h(\tilde{I})$ based on $\tilde{I}$.    \label{line-stage2-start}
    
    \ForEach {$I$ {\rm \textbf{in}} $T_h(\tilde{I})$}{
 
        run $I$ and measure $E(I)$ and $speed(I)$, record them.

        \If{$speed(I) \ge speed(\tilde{I}) \cdot (1-\epsilon)$ }{
            // $I$ violates speed constraint, pop it
        
            $T_h(\tilde{I})$.pop(I) 
        }
    }

    $I^*\longleftarrow \arg\min_{I\in T_h(\tilde{I})} E_h(I)$ \label{line-stage2-end}
    
    \Return{$I^*$}
\end{algorithm}

\begin{figure}[!h]
    \centering
    \vspace{-0.5em}
    \includegraphics[width=0.8\columnwidth]{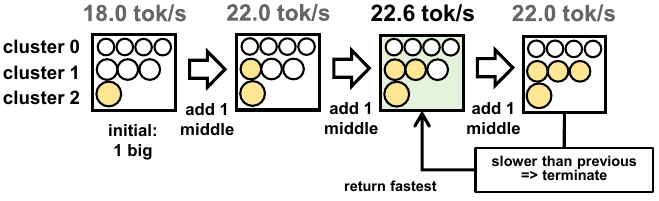}
    \vspace{-0.5em}
    \caption{Stage 1 process on Mate 40 Pro: Starting from 1 big core, middle cores are added sequentially. Stage 1 terminates with selecting 1 big and 2 middle cores.}
    \Description{see main text}
    \label{fig:stage1}
    \vspace{-0.8em}
\end{figure}

The input of stage 1 is the CPU information. Specifically, the max frequency of each CPU cluster is used to distinguish big cores from small ones. The output of this stage is the fastest core selection $\tilde{I}$. The search process starts from 1 prime core, and adds CPU cores greedily from big to small one at a time, meanwhile testing the speed for each core selection plan. Stage 1 terminates if adding 1 more core doesn't speed up any more, or no prime cores and performance cores are left. Efficient cores are not considered. The example search process of stage 1 on Mate 40 Pro is shown in Figure \ref{fig:stage1}.

\textbf{Stage 2: searching for the optimal core selection.}
In stage 2, we search for the optimal core selection $I^*$ of Equation \ref{optimization}. To better carry out our search, we reformulate the original Equation \ref{optimization} into  

\begin{equation}
\begin{aligned}
    &I^* = \arg\min_{I\in S_h(\tilde{I})} E_h(I),\\
    s.t.,&\ \frac{speed(I)}{\max\limits_J speed(J)} \ge 1-\epsilon. 
\end{aligned}
\label{stage2}
\end{equation}
Compared with Equation \ref{optimization}, 1) we change the objective from empirical energy $E(I)$ to heuristic objective $E_h(I)$ by introducing a power heuristic function $h$ to guide the search and enhance searching robustness, and 
2) the search space is scaled down from all core selections $S$ to a candidate subset $S_h(\tilde{I})$ filtered by heuristic function $h$. In what follows, we further introduce the detailed design of power heuristic function $h(I)$, modified objective function $E_h(I)$, and candidate set $S_h(I)$.

\textit{Power heuristic function $h(I)$.} 
The power heuristic function is modeled based on 4 major hardware and OS characteristics, including power-frequency relationship, CPU type modeling, idleness modeling, and CPUFreq governor modeling.
\begin{itemize}
    \item Power-frequency relationship is modeled  quadratically along with a static power $Ps$: $h(I)\propto f^2+Ps$, 
    because recent work \cite{MobiRL} suggests power increases super-linearly corresponding to frequency on modern cores.

    \item CPU type also  determines CPU power. 
    Therefore, each cluster of different CPU type is assigned a distinct scaling factor $a_i$, so that $h(I)\propto \sum_{i=0}^{n-1}a_i\cdot f_i^2+Ps$, where $n$ is the number of CPU clusters. 

    \item Idle cores inside a cluster is assigned a reduced factor $b<1$ based on CPU idle state behaviors \cite{kernel-idle}. Denote the number of cores in cluster $i$ as $|C_i|$ and selected ones as $|I_i|$, the idleness reduced factor of cluster $i$ is summed up to $(|I_i|+(|C_i|-|I_i|)b)$. Up to now, $h(I)$ is formulated as $h(I)\propto\sum_{i=0}^{n-1}a_i(|I_i|+(|C_i|-|I_i|)b) f_i^2 +Ps$.
    
    \item CPUFreq governor behavior is modeled to estimate the operating frequency of cluster $i$ $f_i$. We check the source code\footnote{function named \textit{scale\_load\_to\_cpu} in file kernel/sched/sched.h} in latest Android kernel \cite{sched} and discover that the estimated workload is scaled by the capacity factor $s_I=\frac{selected\ biggest\ capacity}{biggest\ capacity}$, and thus the assigned frequency is usually the max frequency scaled by this factor, $f_i=f_{max,i}\cdot s_I$. 
\end{itemize}

Combining these features, the final power heuristic function is modeled as
\begin{equation}
    h(I)=\sum_{i=0}^{n-1} a_{i} (|I_i|+(|C_i|-|I_i|)b) (f_{max,i}\cdot s_{I})^2 + Ps.
    \label{heuristics}
\end{equation}


\textit{Objective function: $E_h(I)=(1-\alpha)E(I)+\alpha h(I)t(I)$.} The empirically measured energy $E(I)$ suffers from intolerable system fluctuation in energy and speed, as high as $5\%$, which can easily skew the search results. To address this issue, we modify the original objective to be a weighted average of the observed energy $E(I)$ and heuristic estimation $h(I)\cdot t(I)$ to enhance the robustness of the search for the optimal result. Ablation studies in Section \ref{ablation} verifies this design.

\begin{figure}[!h]
    \centering
    \vspace{-0.5em}
    \includegraphics[width=0.6\columnwidth]{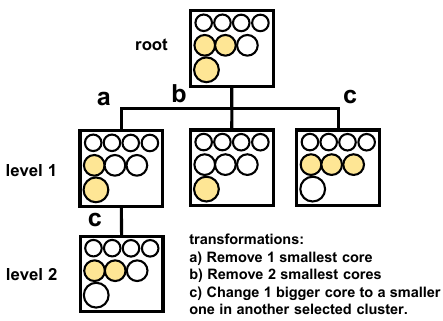}
    \vspace{-0.6em}
    \caption{Stage 2 heuristic tree on Mate 40 Pro: transformations a), b), c) are used to generate nodes.}
    \vspace{-0.7em}
    \Description{see Caption}
    \label{fig:stage2}
\end{figure}

\textit{Generation of candidate set (heuristic tree): $S_h(\tilde{I})$.} To generate competitive candidates, we start from the stage 1 result $\tilde{I}$ as the root and grow a candidate tree. The candidate set $S_h(\tilde{I})$ is defined as all the nodes on the tree. Heuristic transformations are defined to generate succeeding nodes from parents, so that succeeding nodes are more energy-efficient and comparable in speed. On Android, 4 transformations are designed, including a) removing 1 smallest core, b) removing 2 smallest cores, c) changing 1 bigger core to a smaller one in another selected cluster, and d) changing a selected cluster of bigger cores to an unselected cluster of smaller cores. Besides, efficient cores such as Cortex-A5 series are not ignored from selection. The tree depth is limited less than 2, and  transformations a) and b) are only allowed in level 1 to avoid plan duplication. The tree size is therefore capped small. In Figure \ref{fig:stage2}, transformations a), b), c) are applicable to the root, which has 1 big core and 2 middle, generating 3 nodes at level 1. The first node at level 1 can further generate a node in level 2 with c), resulting in a tree with 5 candidates.

\noindent On iOS, the generation process is much easier. Only reducing 1 thread is valid to generate candidate node. 

With the heuristic candidate tree, candidate set is 5-10× smaller while preserving the optimality of the search. A detailed analysis is presented in Section \ref{ablation}.

To effectively solve the optimization defined in Equation \ref{stage2}, stage 2 searching is introduced,  presented in lines \ref{line-stage2-start}-\ref{line-stage2-end} of Algorithm \ref{alg:AECS}. Stage 2 inputs CPU information to compute the power heuristic function, along with the stage 1 result, and outputs the energy-optimal core selection, which is also the final output of entire AECS searching. First, the candidate set $S_h(\tilde{I})$ is generated.
Then, stage 2 searching proceeds over the candidate set $S_h(\tilde{I})$ and  measures the speed and energy of each candidate. If a candidate violates the speed constraint, it's immediately removed from the candidate set.
After all candidates are tested, the algorithm outputs the candidate with minimum value on the modified energy objective.

  


    

            
    
    




\section{Implementation}\label{implementation}

We integrate AECS into the on-device inference engine MNN and implement an energy profiling module to form our version, MNN-AECS. Core functionality is implemented in approximately 6,000 lines of C++, Java, Swift, Objective C, and Python code.

\begin{table}[!h]
    \centering
    \caption{Model compatibility of different engines on Android. \ding{52} denotes support, oom denotes out of memory on some devices, and \ding{53} denotes not supported.}
    \vspace{-0.5em}
    \resizebox{\columnwidth}{!}{
    \begin{tabular}{cccccc}
    \toprule[1.5pt]
          & Qwen2.5-1.5B & Qwen2.5-3B & Llama3.2-1B & Llama3.2-3B & Gemma2-2B \\
        \midrule
        executorch & \ding{53} & \ding{53} & \ding{52} & \ding{52}& \ding{53}\\
        llama.cpp & \ding{52} & \ding{52} & \ding{52} & \ding{52} & \ding{52} \\
        MediaPipe & \ding{53} & \ding{53} & \ding{53} & \ding{53} & \ding{52} \\
        mllm & \ding{52} & oom & \ding{52} & oom & \ding{53}\\
        MNN & \ding{52} & \ding{52} & \ding{52} & \ding{52} & \ding{52} \\
    \bottomrule
    \end{tabular}
    }
    \vspace{1em}
    \label{tab:engine-model}
    \vspace{-2em}
\end{table}

\subsection{Engine Integration (MNN-AECS)}

We choose MNN as our base engine because of its superior model compatibility (Table \ref{tab:engine-model}) and its fast CPU LLM decoding speed. MNN decodes 1.1$\times$ to  3$\times$ faster than nearly all other engines \cite{llama.cpp, executorch, MediaPipe, mllm}. This speed advantage is attributed to its contiguous KV cache and weight layout. 

The integration involves outside-engine AECS search and inside-engine thread pool and memory pool modifications.


\textbf{Outside-engine AECS search.}
AECS searching in Algorithm \ref{alg:AECS} is implemented outside MNN. Algorithm inputs CPU information including CPU capacity, max frequency and cluster information, and CPU type from \textit{/sys/devices/system/cpu/} and \textit{/proc/cpuinfo}. During searching, AECS calls the newly-added core selection interface of modified MNN and profiles energy through energy profiling module. After searching, the optimized core selection is recorded and used in  following MNN LLM decoding.

\textbf{Inside-engine modifications.} The primary requirement for MNN modification is to enable rapid switching between different core selections. This functionality is crucial for AECS searching, which necessitates efficient testing of various core selections. Besides, our design employs distinct core selections for prefill and decode to prevent interference, where rapid switching is essential for inference speed. However, the original MNN doesn't support changing core selection. To address this challenge, we modify MNN thread pool and memory pool.

\textbf{Thread pool modifications.} We utilize Android system call \textit{\_\_NR\_sched\_setaffinity} to set and reset CPU bindings, while exposing an additional thread number reset interface on iOS. Thanks to the modified interfaces, core selection can be changed with a simple function call. 

\textbf{Memory pool modifications.}
Originally, MNN's KV cache memory buffer layout depends on thread number.
We eliminate such dependency in our modified memory pool, so that KV cache can be shared by prefill and decode with different thread number.



\subsection{Energy Profiling Module}
\textbf{Requirements.} Energy profiling module needs to satisfy 3 key requirements: 1)  \textit{Precision}: Energy profiling is required to be highly precise. 
2) \textit{Interoperability with MNN-AECS}: Energy data from OS interface, regardless of the interface language, shall be able to pass to our C++ engine and Python evaluation pipeline. 3) \textit{Without developer mode}: Developer mode is not permitted for a commercial app. However, it can be used during evaluation, as evaluation is not part of the LLM serving app.

\textbf{Challenges.} For Android, energy profiling is only available from Java via  \textit{android.os.BatteryManager}. The OS interface updates every 250 ms, making tests shorter than 500 ms imprecise. Besides, the interface is in Java, but our engine is implemented in C++. 

For iOS, energy can only be monitored on PC through tunneld over WLAN. Furthermore, iOS energy profiling is impossible without developer mode.
Though iOS Sysdiagnose \cite{Sysdiagnose} and Xcode energy gauge\footnote{\textit{com.apple.xcode.debug-gauge-data-providers.Energy}} can both measure energy performance of an iOS app, Sysdiagnose updates energy every 20s and suffers from $\ge10\%$ fluctuation, failing to meet our precision requirement. In contrast, Xcode energy gauge is more precise but necessitates developer mode.

\begin{figure}[!h]
    \centering
    \includegraphics[width=0.75\columnwidth]{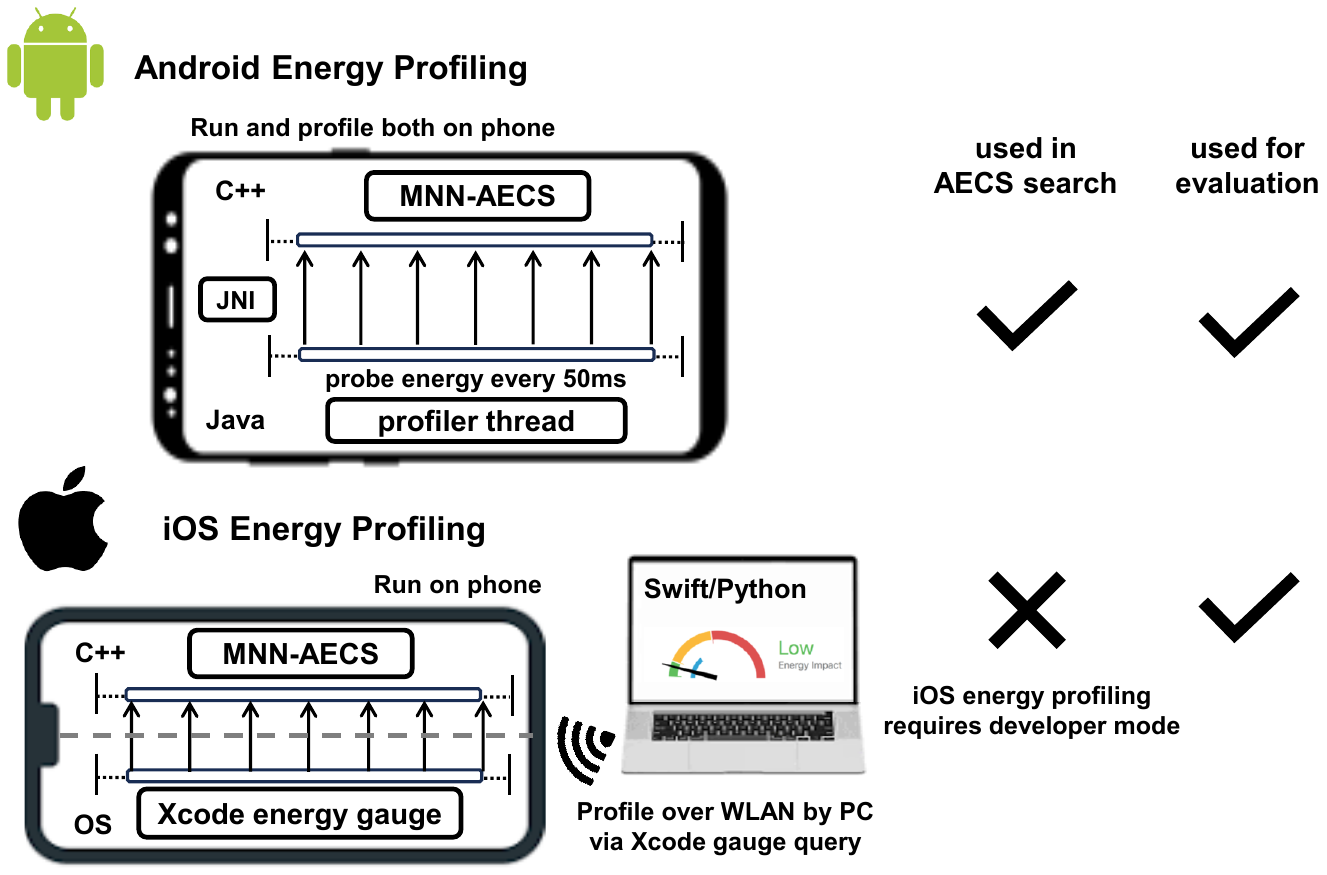}

    \vspace{-0.5em}
    \caption{MNN-AECS energy profiling module.} 
    \label{fig:profile}
    \Description{See main text description}
    \vspace{-1em}
\end{figure}

\textbf{Implementations.}
For Android, we pool current and voltage in a separate profiling thread to calculate energy, which is then passed to MNN-AECS over JNI (Java-Native Interface), illustrated in Figure \ref{fig:profile}. To make the energy profiling more precise, we probe \textit{BatteryManager} every 50 ms so that OS energy updates can be monitored in time. For AECS search, we decode 50 tokens, so that the decode time exceeds the OS interface update interval, ensuring the precision.

For iOS, we opt for Xcode energy gauge to get more precise power data, despite it providing only relative power without exact units. Illustrated in  Figure \ref{fig:profile}, energy of MNN-AECS is monitored by Xcode energy gauge on PC over WLAN though tunneld implemented in \textit{pymobiledevice3} \cite{pymobiledevice3}. AECS search aborts energy profiling on iOS, using only the heuristics for optimization, due to the developer mode requirement of Xcode energy gauge. 
The iOS energy profiling module is utilized exclusively in evaluation experiments, standalone from MNN-AECS.

\section{Evaluation}
\label{evaluation}

\subsection{Experimental Setup}
\label{setup}
We implement a novel cross-platform testbed to support our extensive experiments on both Android and iOS devices, comparing MNN-AECS energy and speed performance with 5 state-of-the-art on-device LLM engines across 5 popular LLMs over 4 real-world datasets.

\textbf{Devices.} To ensure extensive coverage of common-seen mobile devices, 5 different Android phones and 2 iOS phones are used in experiments, listed in Table \ref{tab:devices}.  

\textbf{Baseline engines and LLMs.} On Android, experiments are conducted between our MNN-AECS and its base engine MNN, along with other 4 engines: llama.cpp \cite{llama.cpp}, mllm \cite{mllm}, MediaPipe \cite{MediaPipe}, and executorch \cite{executorch}, over Llama3.2-1B/3B \cite{Llama3.2}, Qwen2.5-1.5B/3B \cite{Qwen2.5}, and Gemma2-2B \cite{Gemma2}. If an engine doesn't support a model, shown in Table \ref{tab:engine-model}, we skip the model test on that engine. Qwen and Llama models are 4-bit quantized and Gemma is 8-bit quantized for fair comparison across engines. On iOS, we compare across ours, MNN, and llama.cpp, over Llama3.2-1B and Qwen2.5-1.5B. 

\textbf{Testing benchmarks.} 2 types of tests are adopted in our analysis. 1) \textbf{Fixed length experiments:} In order to evaluate MNN-AECS performance under different prefill prompt lengths and decode generation lengths, we control LLM inputs and outputs, so that prefill lengths are in $\{64,256,1024\}$, and decode lengths in $\{128, 256, 512\}$. 2) \textbf{Dataset experiments:} To measure the performance of models on practical daily tasks, we conduct dataset experiments on 4 different mobile LLM use cases: math problem solving (MathQA \cite{MathQA}), open domain QA (TruthfulQA \cite{TruthfulQA}), multi-turn conversation (ShareGPT \cite{ShareGPT}), and role play (RoleBench \cite{RoleBench}). A representative subset randomly sampled from each dataset is used in test due to computation and battery constraints on phone.

\textbf{Metrics.} 4 major metrics are compared: energy (mJ/token), battery ($\mu$Ah/token), decode speed (token/s), and CPU use.

\textbf{Testing conditions.} To simulate real-world use condition, we conduct all tests \textbf{unplugged}. Besides, alike previous works \cite{MobiRL}, we cool down the phones to under $40^\circ C$ before each test to prevent high temperature throttling. For those phones where temperature information isn't available, they sleep for about a minute to cool down. The battery charge is also kept above $50\%$ to prevent low-battery throttling.

\subsection{Tuned Results}
The tuned core selections of MNN-AECS on 7 devices are presented in Table \ref{tab:tune_results}. 
These once-and-for-all tuned results are subsequently used across all our experiments. 

\textbf{Lower CPU utilization.} Table \ref{tab:dataset-CPU-use} illustrates the significant CPU utilization reduction of MNN-AECS over the baselines. Other engines utilize $4\sim8$ cores. By contrast, MNN-AECS uses only $\leq 2$ cores across all 7 devices, reducing $50-75\%$ CPU use on almost all devices remarkably. Only llama.cpp on iPhone 15 has comparably low CPU utilization.

\begin{table}[!h]
    \centering
    \vspace{-0.5em}
    \caption{Tuned core selections of AECS on 7 devices. (Refer to Table \ref{tab:devices} for complete device hardware info.)}
    \vspace{-0.5em}
\resizebox{0.8\columnwidth}{!}{
    \begin{tabular}{ccc}
    \toprule[1.5pt]
         &{\bf device} & {\bf tuned core selections} \\
         \midrule
         \multirow{5}{*}{\bf Android}& Mate 40 Pro & {2*A77(2.54GHz)}  \\
         &{Honor V30 Pro} &{2*A76(2.36GHz)} \\
         & Galaxy A56 & {2*A720(2.6GHz)}  \\
         & {Meizu 21}& 1*X4(3.3GHz) +1*A720(2.96GHz) \\
         & {Xiaomi 15 Pro} & {2*Oryon(4.32GHz)} \\
         \midrule
         \multirow{2}{*}{\bf Apple} & iPhone 12 & 1 thread \\
         & iPhone 15 & 2 thread \\
    \bottomrule[1.5pt]
    \end{tabular}
    }    \label{tab:tune_results}
    \vspace{-0.5em}
\end{table}

\begin{table}[!h]
    \centering
    \caption{CPU core utilization in decode phase. MNN-AECS reduces core utilization by $50-75\%$.}
    \vspace{-0.5em}
\resizebox{0.95\columnwidth}{!}{
    \begin{tabular}{cccccccc}
    \toprule[1.5pt]
        \multirow{3}{*}{\bf engine} & \multicolumn{5}{c}{\bf selected core number} & \multicolumn{2}{c}{\bf thread number} \\
        \cmidrule{2-6} \cmidrule{7-8}
        & Mate & V30  & Galaxy & Meizu & Xiaomi & iPhone & iPhone \\
        &  40 Pro & Pro & A56 & 21 & 15 Pro & 12 & 15 \\
         \midrule
        executorch & 8 & 8 & 8 & 8 & 8 & - & - \\
        llama.cpp & 8 & 8 & 8 & 8 & 8 & 2 & \underline{\textbf{2}}\\
        MediaPipe & 4 & 4 & 4 & - & 4 & - & -\\
        mllm & 4 & 4 & 4 & 4 & 4 & - & -  \\
        MNN & 4 & 4 & 4 & 4 & 4 & 4 & 4\\
        MNN-AECS & \underline{\textbf{2}} & \underline{\textbf{2}} & \underline{\textbf{2}} & \underline{\textbf{2}} & \underline{\textbf{2}} & \underline{\textbf{1}} & \underline{\textbf{2}} \\
    \bottomrule[1.5pt]
    \end{tabular}
    }
    \label{tab:dataset-CPU-use}
    \vspace{-1em}
\end{table}




\subsection{Fixed Length Experiments}
\begin{figure}[!h]
    \vspace{-0.5em}
    \centering
    \subfigure[Impact of different prompt and decode lengths on energy.]{\includegraphics[width=0.95\linewidth]{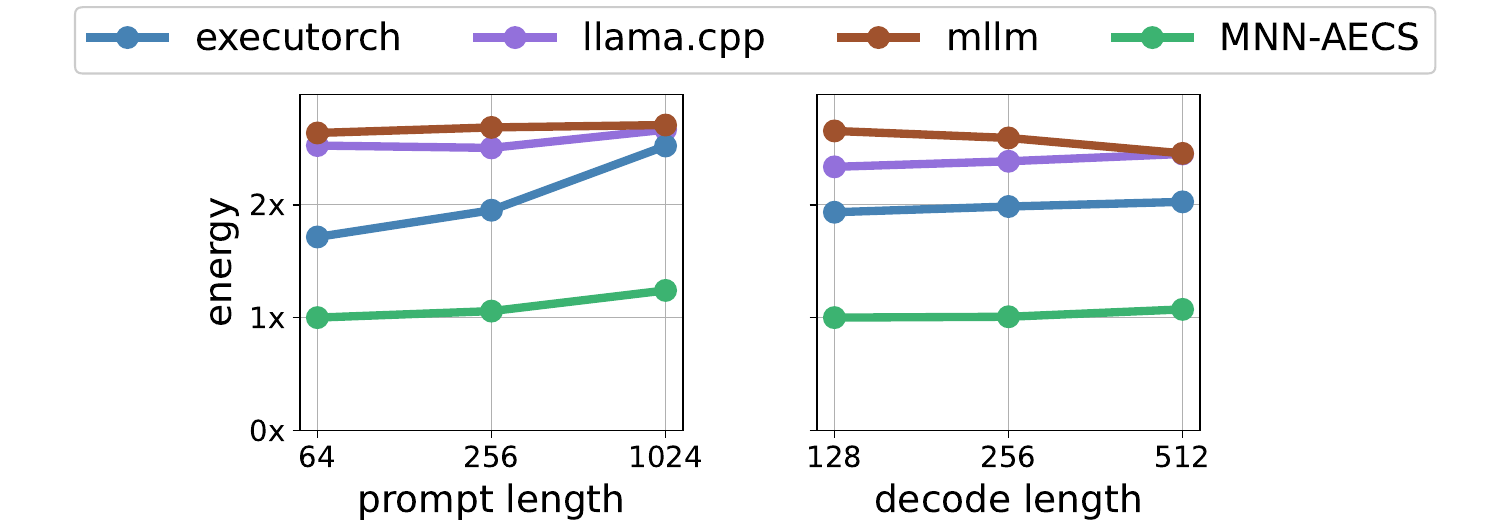}}
    
    \vspace{-1em}
    \subfigure[Impact of different prompt and decode lengths on decode speed.]{\includegraphics[width=0.95\linewidth]{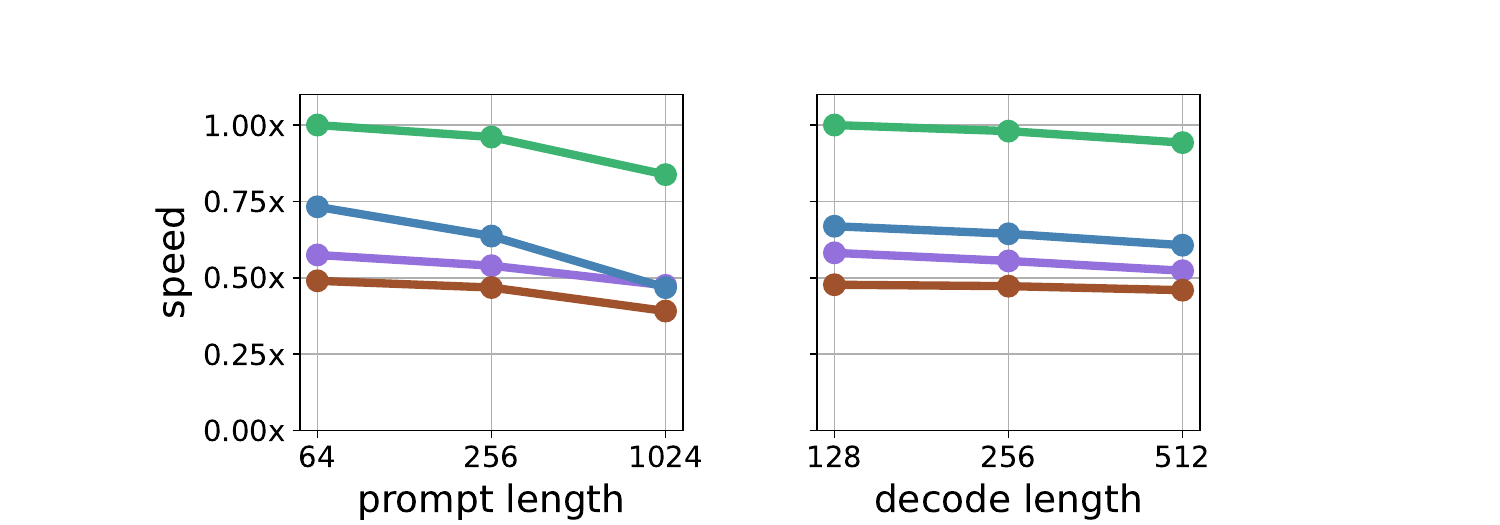}}
    
    \vspace{-0.5em}
    \caption{Normalized energy and decode speed comparison between MNN-AECS and baselines across 7 devices and 5 models.}
    \vspace{-0.6em}
    \label{fig:length-relationship}
    \Description{See main text}
    \vspace{-0.6em}
\end{figure}

\begin{figure}[!h]
    \centering
    \vspace{-0.5em}
    \subfigure[Impact of prompt length on energy.]{\includegraphics[width=\linewidth]{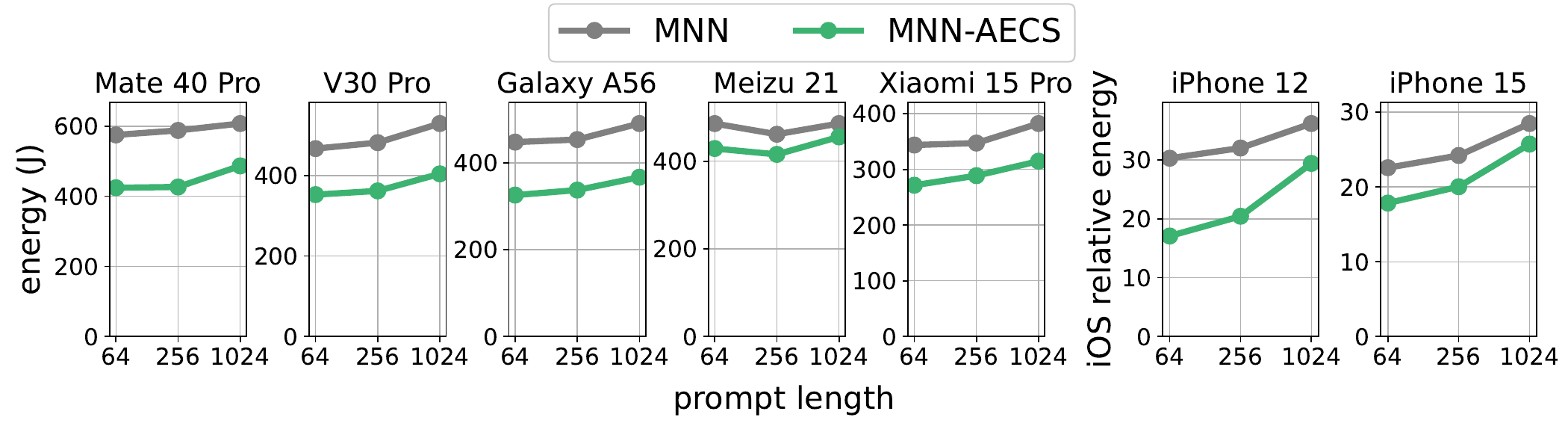}}

    \vspace{-1em}
    \subfigure[Impact of decode length on energy.]{\includegraphics[width=\linewidth]{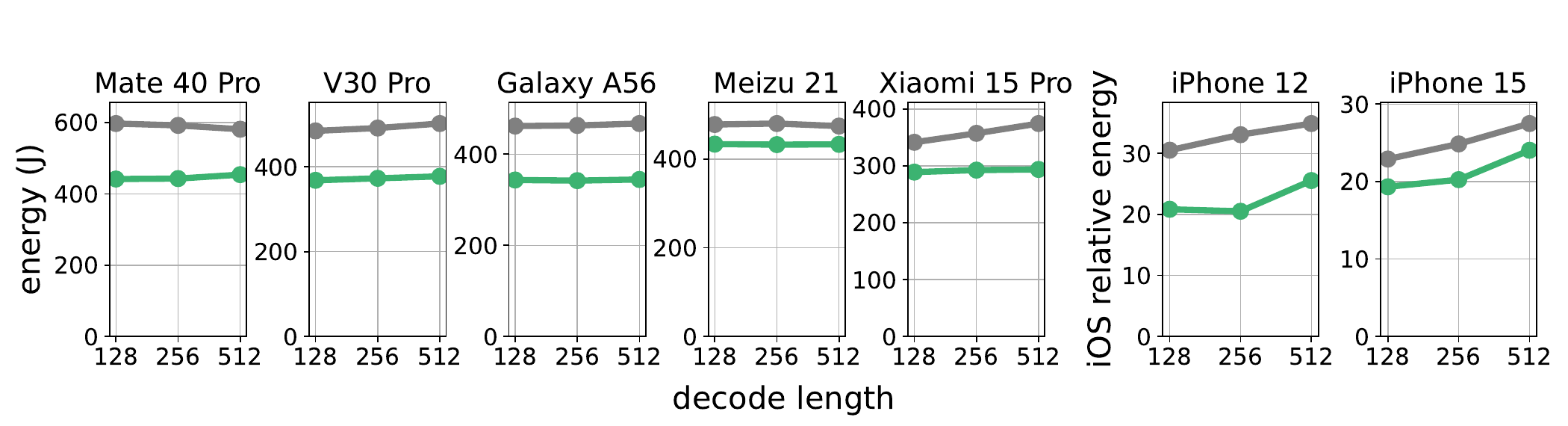}}

    \vspace{-0.5em}
    \caption{MNN-AECS energy reduction over MNN under different (a) prompt and (b) decode length.}
    \vspace{-0.5em}
    \label{fig:length-relationship-MNN-energy}
\end{figure}

\begin{figure}[!h]
    \vspace{-0.5em}
    \centering
    \subfigure[Impact of prompt length on decode speed.]{\includegraphics[width=\linewidth]{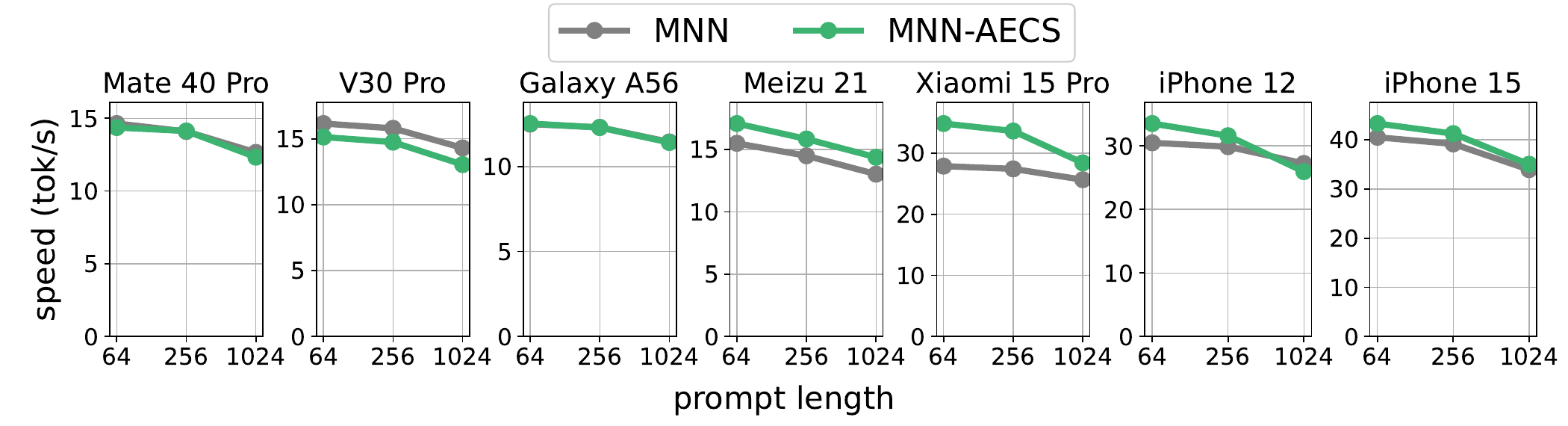}}

    \vspace{-1em}
    \subfigure[Impact of decode length on decode speed.]{\includegraphics[width=\linewidth]{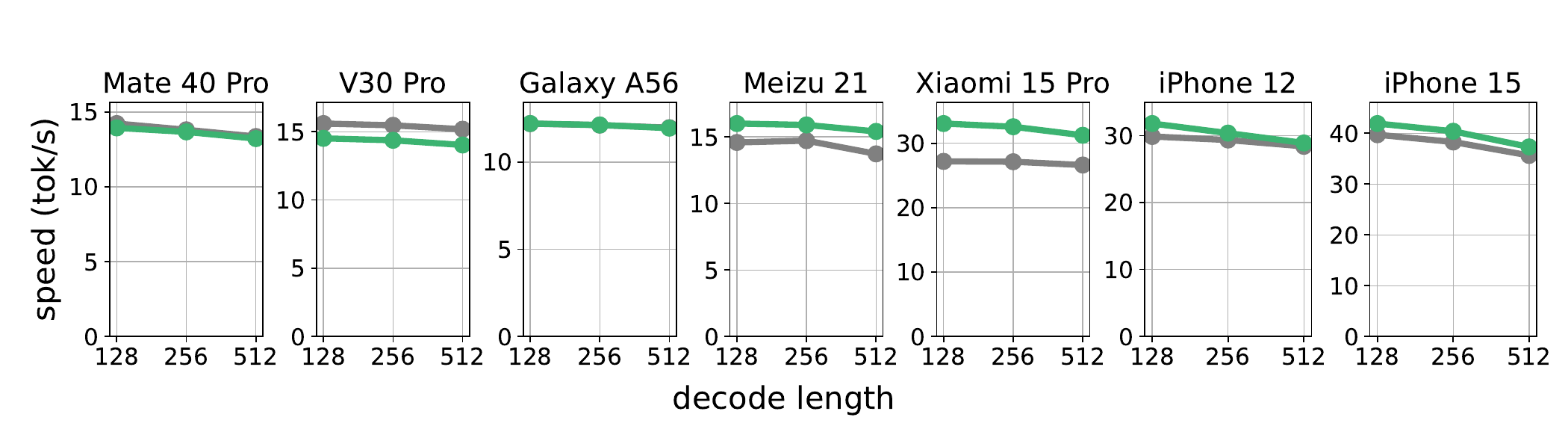}}
    \vspace{-0.8em}
    \caption{MNN-AECS decode speed compared to MNN under different (a) prompt and (b) decode length.}
    \vspace{-0.6em}
    \label{fig:length-relationship-MNN-speed}
\end{figure}

To analyze energy and decode speed of MNN-AECS under different prompt and decode generation lengths, we normalize the geometric mean across all devices and models and compare MNN-AECS with executorch, llama.cpp, and mllm in Figure \ref{fig:length-relationship}. Also, more detailed analyses on each device are provided between MNN-AECS and original MNN in Figures \ref{fig:length-relationship-MNN-energy} and \ref{fig:length-relationship-MNN-speed}.

\begin{table*}[!t]
    \centering
    \caption{Dataset experiments results on 5 Android devices averaged over 4 datasets.}
    \vspace{-0.7em}
\resizebox{0.85\linewidth}{!}{
    \begin{tabular}{ccccccccccccccccc}
    \toprule[1.5pt]
    \multirow{2}{*}{\bf device} & \multirow{2}{*}{\bf engine} & \multicolumn{3}{c}{\bf Qwen2.5-1.5B} & \multicolumn{3}{c}{\bf Qwen2.5-3B} & \multicolumn{3}{c}{\bf Llama3.2-1B} & \multicolumn{3}{c}{\bf Llama3.2-3B} & \multicolumn{3}{c}{\bf Gemma2-2B} \\
    \cmidrule(r){3-5}  \cmidrule(r){6-8}  \cmidrule(r){9-11}  \cmidrule(r){12-14} \cmidrule(r){15-17}
         & & {\bf speed} & {\bf battery} & {\bf energy} & {\bf speed} & {\bf battery} & {\bf energy} & {\bf speed} & {\bf battery} & {\bf energy} & {\bf speed} & {\bf battery} & {\bf energy} & {\bf speed} & {\bf battery} & {\bf energy} \\
        \midrule
    \multirow{6}{*}{Mate 40 Pro} & executorch & - & - & - & - & - & - & 14.4 & 34.0 & 505.6 & 5.6 & 87.6 & 1246.6 & - & - & - \\
	 & llama.cpp & 10.1 & 53.0 & 772.2 & 6.3 & 77.7 & 1143.0 & 15.8 & 35.7 & 509.6 & 5.8 & 85.0 & 1223.7 & - & - & - \\
	 & MediaPipe & - & - & - & - & - & - & - & - & - & - & - & - & 3.1 & 372.5 & 5437.9 \\
	 & mllm & - & - & - & - & - & - & 7.3 & 73.0 & 1080.2 & - & - & - & - & - & - \\
	 & MNN & \underline{\textbf{20.7}} & 26.8 & 389.4 & 10.5 & 50.0 & 737.7 & \underline{\textbf{26.2}} & 21.2 & 314.6 & \underline{\textbf{9.7}} & 54.1 & 794.9 & \underline{\textbf{12.3}} & 44.1 & 662.2 \\
	 & MNN-AECS & 20.3 & \underline{\textbf{20.2}} & \underline{\textbf{298.2}} & \underline{\textbf{10.5}} & \underline{\textbf{39.3}} & \underline{\textbf{583.6}} & 25.8 & \underline{\textbf{16.4}} & \underline{\textbf{236.7}} & 9.4 & \underline{\textbf{45.5}} & \underline{\textbf{650.3}} & 12.2 & \underline{\textbf{36.4}} & \underline{\textbf{523.1}} \\
	\midrule
	\multirow{6}{*}{V30 Pro} & executorch & - & - & - & - & - & - & 16.9 & 28.6 & 410.8 & 7.2 & 65.6 & 955.1 & - & - & - \\
	 & llama.cpp & 9.9 & 50.3 & 728.1 & 6.3 & 73.9 & 1088.6 & 15.7 & 33.3 & 467.0 & 6.0 & 82.0 & 1166.0 & - & - & - \\
	 & MediaPipe & - & - & - & - & - & - & - & - & - & - & - & - & 4.7 & 96.7 & 1418.7 \\
	 & mllm & - & - & - & - & - & - & 7.7 & 64.3 & 926.8 & - & - & - & - & - & - \\
	 & MNN & \underline{\textbf{23.0}} & 21.0 & 313.0 & \underline{\textbf{11.7}} & 43.0 & 619.7 & \underline{\textbf{29.0}} & 17.3 & 245.7 & \underline{\textbf{10.9}} & 48.3 & 675.7 & \underline{\textbf{14.0}} & 38.1 & 541.7 \\
	 & MNN-AECS & 20.9 & \underline{\textbf{17.2}} & \underline{\textbf{250.3}} & 11.1 & \underline{\textbf{32.8}} & \underline{\textbf{479.9}} & 26.4 & \underline{\textbf{13.4}} & \underline{\textbf{199.8}} & 10.3 & \underline{\textbf{36.7}} & \underline{\textbf{525.2}} & 13.3 & \underline{\textbf{27.7}} & \underline{\textbf{400.2}} \\
	\midrule
	\multirow{6}{*}{Galaxy A56} & executorch & - & - & - & - & - & - & 17.6 & 24.2 & 356.8 & 7.3 & 58.0 & 846.9 & - & - & - \\
	 & llama.cpp & 10.8 & 44.8 & 661.4 & 7.2 & 62.1 & 920.5 & 17.5 & 29.5 & 422.4 & \underline{\textbf{9.1}} & 60.6 & 871.7 & - & - & - \\
	 & MediaPipe & - & - & - & - & - & - & - & - & - & - & - & - & 4.3 & 59.4 & 916.4 \\
	 & mllm & - & - & - & - & - & - & 6.9 & 45.7 & 669.0 & - & - & - & - & - & - \\
	 & MNN & \underline{\textbf{18.3}} & 18.9 & 290.8 & 9.0 & 33.2 & 501.1 & 21.8 & 16.1 & 235.8 & 8.4 & 36.6 & 560.9 & 10.8 & 32.7 & 492.4 \\
	 & MNN-AECS & 18.0 & \underline{\textbf{16.1}} & \underline{\textbf{237.2}} & \underline{\textbf{9.2}} & \underline{\textbf{28.3}} & \underline{\textbf{429.5}} & \underline{\textbf{22.2}} & \underline{\textbf{12.0}} & \underline{\textbf{185.0}} & 8.4 & \underline{\textbf{31.5}} & \underline{\textbf{469.5}} & \underline{\textbf{11.0}} & \underline{\textbf{25.3}} & \underline{\textbf{375.0}} \\
	\midrule
	\multirow{5}{*}{Meizu 21} & executorch & - & - & - & - & - & - & 22.4 & 25.4 & 368.8 & 9.5 & 57.4 & 841.6 & - & - & - \\
	 & llama.cpp & 11.3 & 56.9 & 853.4 & 7.8 & 81.3 & 1179.0 & 18.2 & 36.6 & 525.3 & 7.3 & 83.2 & 1226.8 & - & - & - \\
	 & mllm & - & - & - & - & - & - & 9.6 & 44.1 & 674.4 & - & - & - & - & - & - \\
	 & MNN & 21.8 & 19.2 & 283.6 & 11.6 & 35.9 & 548.6 & 26.8 & 14.8 & 228.6 & 10.3 & 39.9 & 585.2 & 11.1 & 37.2 & 552.0 \\
	 & MNN-AECS & \underline{\textbf{24.4}} & \underline{\textbf{18.1}} & \underline{\textbf{262.3}} & \underline{\textbf{12.9}} & \underline{\textbf{31.4}} & \underline{\textbf{472.6}} & \underline{\textbf{30.1}} & \underline{\textbf{13.9}} & \underline{\textbf{213.0}} & \underline{\textbf{11.4}} & \underline{\textbf{35.9}} & \underline{\textbf{513.2}} & \underline{\textbf{12.6}} & \underline{\textbf{35.0}} & \underline{\textbf{509.0}} \\
	\midrule
	\multirow{6}{*}{Xiaomi 15 Pro} & executorch & - & - & - & - & - & - & 39.8 & 13.8 & 195.3 & 21.1 & 33.1 & 480.3 & - & - & - \\
	 & llama.cpp & 33.5 & 21.2 & 318.7 & 14.1 & 50.2 & 756.8 & 41.5 & 18.1 & 267.8 & 15.1 & 50.1 & 742.2 & - & - & - \\
	 & MediaPipe & - & - & - & - & - & - & - & - & - & - & - & - & 12.6 & 46.1 & 692.5 \\
	 & mllm & - & - & - & - & - & - & 15.2 & 38.1 & 558.6 & - & - & - & - & - & - \\
	 & MNN & 37.1 & 14.2 & 213.9 & 20.1 & 28.8 & 416.4 & 47.2 & 9.8 & 143.5 & 19.0 & 29.7 & 443.0 & 23.4 & 25.4 & 359.4 \\
	 & MNN-AECS & \underline{\textbf{45.6}} & \underline{\textbf{9.9}} & \underline{\textbf{150.4}} & \underline{\textbf{23.3}} & \underline{\textbf{22.1}} & \underline{\textbf{322.3}} & \underline{\textbf{59.0}} & \underline{\textbf{7.9}} & \underline{\textbf{114.3}} & \underline{\textbf{21.3}} & \underline{\textbf{22.9}} & \underline{\textbf{355.6}} & \underline{\textbf{28.7}} & \underline{\textbf{20.8}} & \underline{\textbf{292.6}} \\
    \bottomrule[1.5pt]
    \end{tabular}
    }
    \label{tab:android-dataset}
    \vspace{-0.3em}
\end{table*}

Illustrated in Figure \ref{fig:length-relationship}, for energy, 
MNN-AECS achieves 36\%-62\%, 49\%-60\%, 52\%-54\% energy reduction under prompt length = 64, 256, 1024, and 49\%-61\%, 50\%-61\%, 50\%-56\% energy reduction under decode length = 128, 256, 512. The results indicate that MNN-AECS energy reduction is consistent under different LLM input and output.
For decode speed, MNN-AECS achieves 33\%-104\%, 58\%-104\%, 69\%-113\% speedup under prompt length = 64, 256, 1024, and 43\%-110\%, 45\%-107\%, 45\%-104\% speedup under decode length = 128, 256, 512. The results indicate that MNN-AECS not only reduces the energy but also excels in decode speed.

Figures \ref{fig:length-relationship-MNN-energy} and \ref{fig:length-relationship-MNN-speed} compare MNN-AECS with its base engine  MNN on 7 devices. Energy reduction of MNN-AECS is more significant for shorter prompts, where decode energy makes up for a higher proportion of total energy. Decode length has less impact on energy by contrast. MNN-AECS reduces 18\% to 42\% energy except for Meizu 21 (10\%), whose OS does not scale down the CPU cluster frequency though idle making our strategy less effective.

For decode speed, MNN-AECS only slows down for 6\% on V30 Pro and speeds up 0 to 20\% on all the rest, thanks to MNN-AECS's core affinity and less cores which contributes to less congestion and conflicts on bus. Prompt length also impacts more on decode speed, because longer prompts makes up large KV matrices for decode.



\subsection{Dataset Experiments}
Tables \ref{tab:android-dataset} and \ref{tab:ios-dataset} show the dataset experiments results on 5 Android and 2 iOS devices,  averaged over 4 real-world datasets specified in Section \ref{setup}. We consumes the lowest energy and battery across all devices and models compared to our baselines. MNN-AECS is also the fastest solution in most cases, where slowdown is less than 7\% in the rest.

\begin{figure*}[ht!]
    \centering
    \subfigure[MNN-AECS energy reduction is presented atop subfigures (26\%-92\%).]{\includegraphics[width=0.9\linewidth]{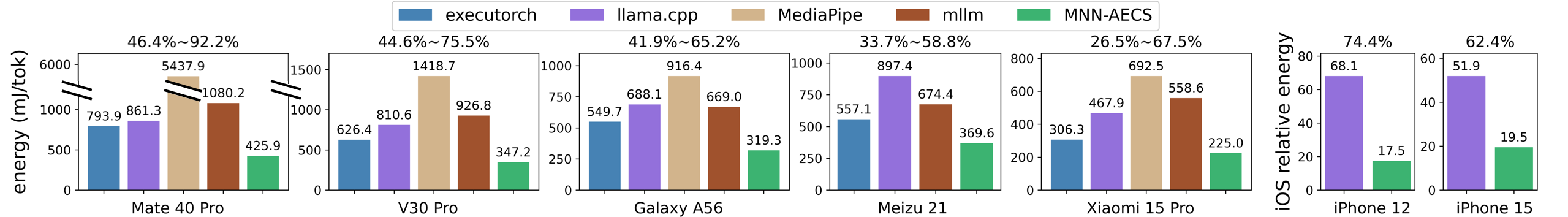}}

    \vspace{-1em}
    \subfigure[MNN-AECS speedup is presented atop subfigures (12\%-363\%).]{\includegraphics[width=0.9\linewidth]{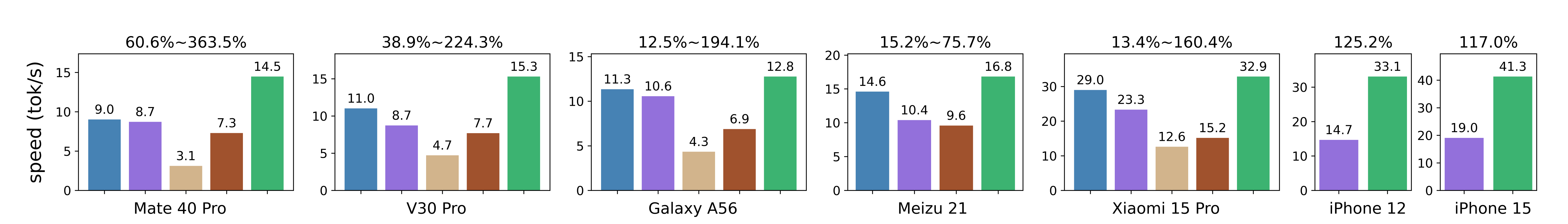}}
    \vspace{-1em}
    \caption{Energy and decode speed in dataset experiments, geometrically averaged across 5 LLMs on 7 devices.}
    \label{fig:dataset-summary}
    \Description{see main text}
    \vspace{-0.6em}
\end{figure*}

Figure \ref{fig:dataset-summary} compares MNN-AECS with executorch, llama.cpp, MediaPipe, and mllm on each devices, where energy and decode speed are geometric mean over all models and datasets tested on the device. MNN-AECS consistently reduces energy by 26\% to 92\% across all devices, and also speeds up by 12\% to 363\%.

\begin{table}[!h]
    \centering
    \vspace{-0.5em}
    \caption{Dataset experiments results on 2 iPhones averaged over 4 datasets.}
    \vspace{-0.6em}
\resizebox{0.75\linewidth}{!}{
    \begin{tabular}{cccccc}
    \toprule[1.5pt]
    \multirow{2}{*}{\bf device} & \multirow{2}{*}{\bf engine} & \multicolumn{2}{c}{\bf Qwen2.5-1.5B} & \multicolumn{2}{c}{\bf Llama3.2-1B} \\
    \cmidrule(r){3-4}  \cmidrule(r){5-6}
         & & {\bf speed} & {\bf energy} & {\bf speed} & {\bf energy} \\
    \midrule
    \multirow{3}{*}{iPhone 12} & llama.cpp & 13.4 & 74.3 & 16.0 & 62.4 \\
	 & MNN & 27.9 & 34.0 & 34.5 & 27.1 \\
	 & ours & \underline{\textbf{29.3}} & \underline{\textbf{19.6}} & \underline{\textbf{37.3}} & \underline{\textbf{15.5}} \\
	\midrule
	\multirow{3}{*}{iPhone 15} & llama.cpp & 17.7 & 55.1 & 20.5 & 48.8 \\
	 & MNN & 33.4 & 28.6 & 46.1 & 19.9 \\
	 & ours & \underline{\textbf{35.5}} & \underline{\textbf{23.4}} & \underline{\textbf{48.1}} & \underline{\textbf{16.3}} \\
    \bottomrule[1.5pt]
    \end{tabular}
    }
    \label{tab:ios-dataset}
    \vspace{-0.6em}
\end{table}

\begin{figure}[!h]
    \centering
    \subfigure[MNN-AECS energy reduction over MNN is presented atop subfigures (10\%-42\%).]{\includegraphics[width=\linewidth]{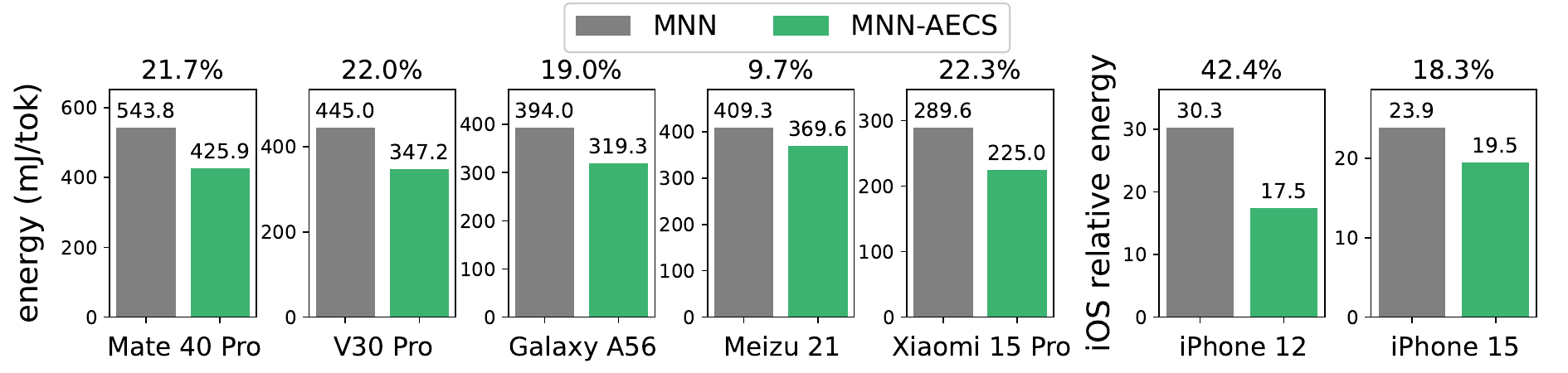}}

    \vspace{-1em}
    \subfigure[MNN-AECS speedup over MNN is presented atop subfigures (-7\%-20\%).]{\includegraphics[width=\linewidth]{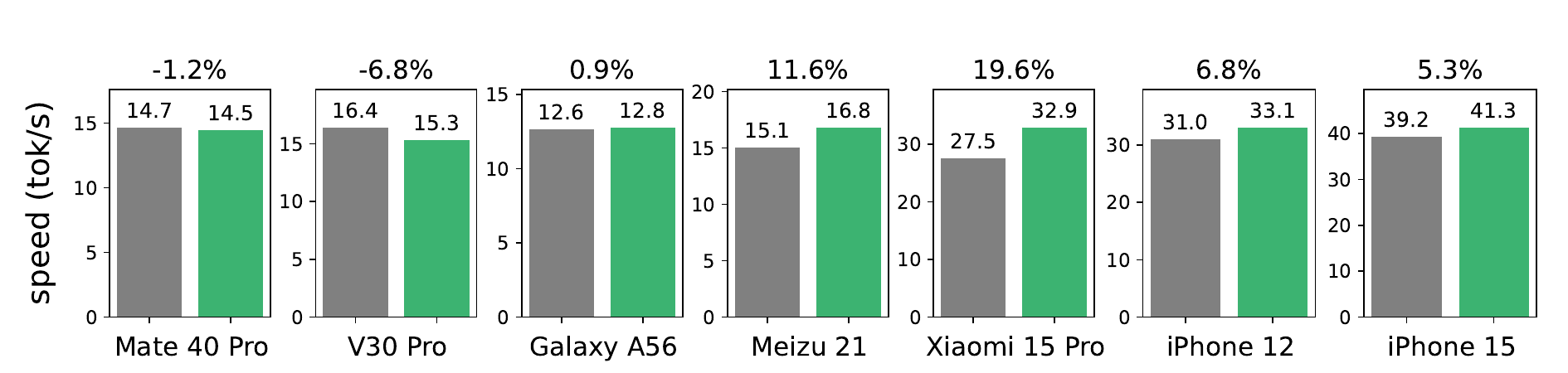}}
    \vspace{-1em}
    \caption{(a) Energy and (b) decode speed comparison between MNN-AECS and MNN across 7 devices in dataset experiments.}
    \vspace{-0.7em}
    \label{fig:MNN-dataset-device}
\end{figure}

\begin{figure}[!h]
    \centering
    \subfigure[MNN-AECS relative energy reduction over MNN on 5 LLMs.]{\includegraphics[width=0.75\linewidth]{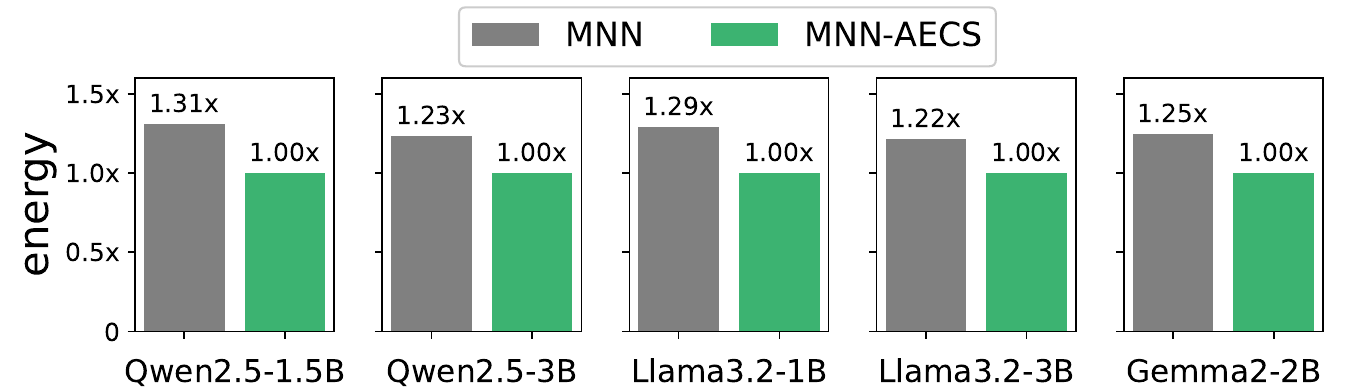}}

    \vspace{-1em}
    \subfigure[MNN-AECS relative speedup over MNN on 5 LLMs.]{\includegraphics[width=0.75\linewidth]{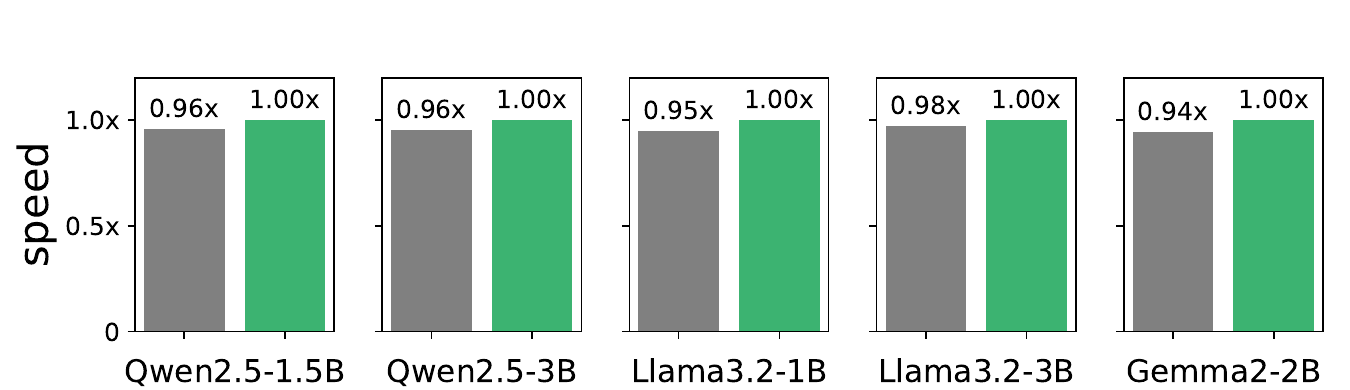}}
    \vspace{-0.5em}
    \caption{(a) Energy and (b) decode speed comparison between MNN-AECS and MNN across 5 models in dataset experiments.}
    \label{fig:MNN-dataset-model}
\end{figure}

To boil down our energy reduction and speedup, we again compare MNN-AECS to original MNN. In Figure \ref{fig:MNN-dataset-device}, we achieve 42\% energy reduction on iPhone 12, about 20\% reduction on 5 other phones, and 10\% reduction on Meizu 21. Only 1\% and 7\% slowdown is witnessed on 2 Huawei phones, and we even speed up 1\% to 20\% on the rest 5. On the other hand, Figure \ref{fig:MNN-dataset-model} shows that MNN-AECS energy reduction and speedup over MNN are consistent across 5 models.


\subsection{Ablation Studies}
\label{ablation}
We compare AECS to exhaustive traversal of all possible core selections to demonstrate optimality and efficiency. Besides, we also analyze how our heuristic-averaged energy objective enhances robustness. Results are shown in Table \ref{tab:ablation}.

\textbf{Optimality and Efficiency: compare to exhaustive traversal.} We conduct exhaustive traversal on all test devices and observe that the searched results are exactly the same as AECS results, which means the optimality rate of AECS is 100\%. The search space of exhaustive traversal scales up to 20-71, taking up 10-20 minutes of foreground execution, which is intolerable for users. AECS successfully removes uncompetitive candidates so that the search time is reduced to 1-2 minutes. The results indicate that AECS has the same optimality rate as exhaustive search while being $10\times$ faster.

\begin{table}[!h]
    \centering
    \caption{Ablation studies: compare AECS to exhaustive traversal and AECS without heuristic.}
    \vspace{-1em}
\resizebox{0.8\columnwidth}{!}{
    \begin{tabular}{cccc}
    \toprule[1.5pt]
         & exhaustive & AECS & \multirow{2}{*}{\bf AECS} \\
        & (optimal) & w/o heuristic &  \\
        \midrule
        search space & 20-71 & 4-9 & 4-9 \\
        serach time (min) & 10-20 & 1-2 & 1-2 \\
        optimality rate & 100\% & 60\%-90\% & 100\% \\
    \bottomrule[1.5pt]
    \end{tabular}
    }
    \label{tab:ablation}
    \vspace{-0.5em}
\end{table}

\textbf{Robustness: compare to AECS without power heuristic.}
In AECS design, the energy objective is a weighted average of measured energy and the power heuristic. Table \ref{tab:ablation} shows that removing the power heuristic leads to fluctuation in searching and consequently lower optimality rate.



\section{Related Works}
\subsection{On-device LLM Inference Engine}
A series of well-known mobile LLM inference engines are available these days, including MNN \cite{Walle}, llama.cpp \cite{llama.cpp}, executorch \cite{executorch}, mllm \cite{mllm}, MediaPipe \cite{MediaPipe} (based on LiteRT \cite{LiteRT}), and MLC-LLM \cite{MLC-LLM} (based on TVM \cite{TVM}). Among these, MNN and llama.cpp support the widest range of models. MNN achieves leading inference speed on both mobile CPUs and GPUs with its optimized KV cache and weights layout that ensures contiguous memory reads. llama.cpp is highly extensible and thus also the most popular. executorch integrates SpinQuant in its engine to improve accuracy, though only supporting Llama series. mllm leverages NPU GEMM kernels to accelerate prefill. MediaPipe and MLC-LLM achieve leading GPU prefill performance through their optimized GPU GEMM kernels. Additionally, PowerInfer-2 \cite{PowerInfer-2} , though not accessible yet, also leverages NPU and activation sparsity to accelerate and accommodate very large LLMs on mobile devices.

These engines primarily focus on prefill acceleration via optimized GEMM kernels, accuracy improvements through quantization, and memory reduction with sparsity. However, none of them address the LLM decoding energy challenge. 

\subsection{Mobile OS DVFS for DL Tasks}
Several mobile DVFS algorithms are designed for DL tasks, such as Asymo \cite{Asymo} and GearDVFS \cite{GearDVFS}. These methods directly manipulate CPU frequency, which  requires root access or customized Android kernel. While they are valuable for OS-level design, they are impractical for engine-level design in our context. Moreover, they specifically target at traditional CNN-based DL tasks, which have significantly different computational characteristics compared to LLM.

\subsection{LLM Quantization and Sparsity}
Quantization and sparsity are 2 major algorithmic methods for LLM speedup and memory reduction. In practice, quantization algorithm, such as k-quant \cite{llama.cpp}, SmoothQuant \cite{SmoothQuant}, SpinQuant \cite{SpinQuant}, AWQ \cite{AWQ}, are commonly adopted to quantize models to 4/8-bit with tolerable accuracy loss to fit the model into the limited memory on mobile devices. On the other hand, sparsity in activation \cite{PowerInfer-2, LLM-in-a-Flash}, KV cache \cite{StreamingLLM}, and MoE structure \cite{EdgeMoE} are also promising directions for reducing memory and computation of LLM inference. Both LLM quantization and sparsity are orthogonal to our core selection system design of core selection.

\section{Discussion on Extensions to Mobile XPU}
\label{discussion}

During memory-bound LLM decoding, XPUs frequently wait for the bus, resulting in limited performance gains from increasing parallelism due to the shared memory bus bottleneck. MNN-AECS exploits such characteristics on CPU to save energy under speed guarantee. Such abstraction and system design can potentially be extended to mobile GPU/NPU.

Latest Snapdragon Adreno GPU automatically reduces shader processors (SP) frequency and mainly leveraging texture processors (TP) during image read \cite{Snapdragon}, which is an idea similar to ours and can be utilized to save energy during GPU LLM decoding. Our preliminary trials to utilize such features in MNN successfully reduce GPU main frequency on Xiaomi 15 Pro (Snapdragon 8 Elite). Though memory bus peaks all the time, ALUs in SP are mostly idle and thus GPU utilization and frequency keep low. On the other hand, existing mobile NPUs primarily aim at GEMM acceleration. LLM decoding can be more energy-efficient, if a hardware API is available to designate a smaller core with less ALUs and registers to process memory-bound Vec-Mat Multiplication.

\section{Conclusion}
We have presented MNN-AECS, a novel engine-level system design for energy-efficient LLM decoding on mobile devices. By adaptively selecting CPU cores, the most energy-efficient core selection is identified for LLM decoding without compromising speed. MNN-AECS has achieved 23\% energy reduction without slowdown compared to original MNN and delivered 39\%–78\% energy reduction and 12\%-363\% decode speedup over 4 other popular on-device LLM engines.

\bibliographystyle{ACM-Reference-Format}
\bibliography{ref}

\end{document}


\maketitle

\section{Magnitude of LLM Decoding Current, Power, and Temperature}
In order to supplement the energy consumption magnitude in the main article, we test and list example magnitude of current, power and temperature in Table \ref{tab:energy-challenge2} for reference.

\section{Theoretical Analysis of Power Heuristic Effectiveness}
The following 2 theorems indicates that as long as the magnitude relationship of h(I) is the same as that of power, the accuracy of the final result will be improved, and the variance will be reduced. We denote $energy(I)$ as $e(I)$, and assume observed test profile $e(I)$ is an unbiased observation of ideal energy function $E(I)$.

\begin{theorem}
\label{th:heuristics-acc}

Higher probability of outputting the optimal result.

\noindent If \ $\forall I,J:E(I) > E(J)\iff h(I) > h(J)$,

\noindent then,
\begin{equation*}
\small
    \begin{aligned}
        &P[obj(I)>obj(J)\ |\  E(I)>E(J)]\\
        =&P[e(I)-e(J)>-\frac{1-\alpha}{\alpha}(h(I)-h(J))\ |\  E(I)>E(J)]\\
        >&P[e(I)>e(J)\ |\ E(I)>E(J)]
    \end{aligned}
\end{equation*}

\noindent The weighted averaging objective $obj(I)=(1-\alpha)e(I) + \alpha h(I)t(I)$ leads to higher accuracy than plain $e(I)$ objective.
\end{theorem}

\begin{theorem}
\label{th:heuristics-var}

Variance Reduction.

Since $e(I)$ is measured by $power(I)\cdot t(I)$ in practice, and these 2 terms are independent, we have the following derivation.

\begin{equation*}
\small
    \begin{aligned}
        &Var[(1-\alpha)e(I)+\alpha h(I)t(I)]\\
        =&Var[(1-\alpha)power(I)+\alpha h(I)]\cdot Var[t(I)]\\
        =&(1-\alpha)^2 Var[e(I)] < Var[e(I)]
    \end{aligned}
\end{equation*}
\end{theorem}

\section{Details of Evaluation Platform}

Our evaluation platform: LLM-Profiler features 2 major characteristics, engine extensibility and comprehensive measurement supports.

\textbf{Engine extensibility.} The system currently supports multiple inference engines such as MNN, llama.cpp, mllm, MediaPipe, and executorch. To resolve symbol conflicts such as the one between llama.cpp and mllm, an engine wrapper is abstracted atop all engines, dynamically loading engine libraries during runtime. Any new engines can be integrated by such wrapping interface.

\begin{table}[!t]
    \centering
    \resizebox{0.75\columnwidth}{!}{
    \begin{tabular}{cccc}
    \toprule[1.5pt]
        \multirow{2}{*}{\bf device} & \multicolumn{3}{c}{\bf peak}  \\
        \cmidrule(r){2-4}
        & {\bf current} & {\bf power} & {\bf temperature} \\
        \midrule
        Xiaomi 15 Pro  &  2386 mA & 9.9 W & $80^\circ C$\\
        Mate 40 Pro & 2189 mA & 8.7 W & $85^\circ C$ \\
    \bottomrule[1.5pt]
    \end{tabular}
    }
    \caption{Peak current, power and CPU temperature of Qwen2.5-1.5B (4-bit quantization), running 20 data entries of multi-turn conversation sampled from ShareGPT dataset, with inference engine MNN.}
    \label{tab:energy-challenge2}
    \Description{See Caption}
\end{table}

\begin{figure}[!t]
    \centering
    \includegraphics[width=\linewidth]{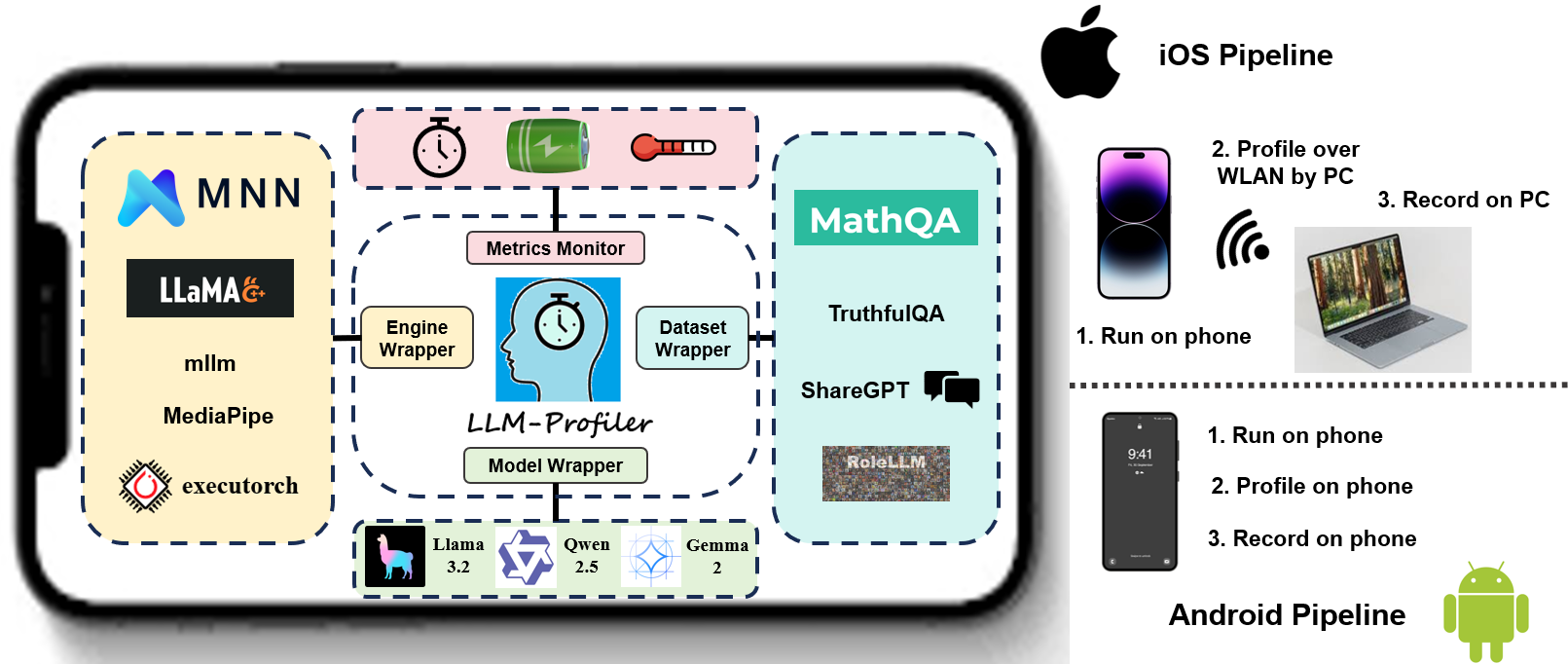}
    \caption{Overview of our evaluation platform \textit{LLM-Profiler}}
    \label{fig:LLM-Profiler}
    \Description{Please see caption.}
\end{figure}

\textbf{Android measurements pipeline.} The whole Android measurements pipeline resides on the phone (Figure \ref{fig:LLM-Profiler}). Speed, energy consumption (current and voltage measurements), battery use (multiplying time and current), and temperature (read from \textit{/sys/class/thermal/thermal\_zone}) are profiled separately for prefill and decode phase. Electricity current data are polled with a time interval of 0.05s, though their values update every 0.25s (1s for Samsung phone) by OS.

\textbf{iOS measurements pipeline.} iOS running and measurements are separated on phone and PC, communicating through tunneld over WLAN (Figure \ref{fig:LLM-Profiler}). Though Sysdiagnose and Xcode energy gauge can both measure energy performance of an iOS app from PC, Sysdiagnose updates energy every 20s and suffers from $\ge10\%$ fluctuation. Therefore, we select Xcode energy gauge \textit{com.apple.xcode.debug-gauge-data-providers.Energy} to get instantaneous power data per second, though such data is only relative power without unit. Besides, CPU utilization rate is also recorded. The pipeline is built upon \textit{pymobiledevice3}.

\section{Experimental Supplementary Details}

\subsection{Engines Versions and Quantization Specification}

\textbf{Engine Versions.} All the engine versions are latest up to February, 2025. We integrate our EECCS into MNN-3.0.4. 3.0.4 is also the version of MNN we compared with. llama.cpp version is b4735, commit hash \textit{73e2ed3c}. mllm version is commit hash \textit{bbf87ffb}. MediaPipe version is 0.10.20. executorch version is 0.5.0, commit hash \textit{e433e610}.

\noindent \textbf{Quantization Specifications.} Because quantization methods impact LLM decoding speed, we carefully choose similar quantization supported across engines to gather fair experimental results.

For MNN, we choose the default quantization: block quantization with block size=128. Qwen-2.5 and Llama-3.2 are quantized into 4 bits, and Gemma-2-2B is quantized into 8 bits to align with MediaPipe.

For llama.cpp, we choose asymmetric 4-bit quantization (Q4\_1) for Qwen-2.5 and Llama-3.2 for fairness, because Q4\_1 has lower computation complexity than MNN block quant. We didn't adopt k-quant because MNN doesn't support super-block, so using it is unfair to llama.cpp.

For mllm, we choose symmetric 4-bit quantization (Q4\_0) for Qwen-2.5 and Llama-3.2 for fairness. mllm only supports Q4\_0 and Q4\_K, so we adopt Q4\_0 for fairness.

For MediaPipe, only asymmetric 8-bit quantized (Q8\_1) Gemma-2-2B model is available.

For executorch, only SpinQuant version Llama-3.2-1B/3B is officially available. We tried Q4\_1 quantization but the models output erroneous texts. Therefore, we use the 4-bit SpinQuant version for comparison. 

\subsection{Statistics of Our Dataset Subsets}
Each subset contains 20 representative samples randomly sampled from the original dataset.

\subsection{Hyper Parameters of EECCS}
\textbf{Slowdown tolerance $\epsilon$.}
We empirically select tolerance $\epsilon=8\%$, which is a slowdown not noticeable to smartphone users, which means the solution is at most 8\% slower than the fastest core selection.

\textbf{decode tuning times.}
The default tuning times in EECCS is 50. Reducing tuning times to less than 50 would lead to lower searching accuracy and suboptimal search results.

\begin{table}[!t]
    \centering
    \caption[Speed, CPU use, power, and energy of Qwen2.5-1.5B prefill and decode on 4 datasets on Xiaomi 15 Pro.]{Speed, CPU use, power, and energy of Qwen2.5-1.5B prefill and decode on 4 datasets on Xiaomi 15 Pro.}
    \resizebox{\columnwidth}{!}{
    \begin{tabular}{cccccc|c}
    \toprule[1.5pt]
        & & \multirow{2}{*}{\textbf{MathQA}} & \multirow{2}{*}{\textbf{TruthfulQA}} & \multirow{2}{*}{\textbf{ShareGPT}} & \multirow{2}{*}{\textbf{RoleBench}} & {\bf ratio} \\
        & & & & & &  \textbf{(D/P)}\\
    \midrule
         \multirow{4}{*}{\bf Prefill} & len & 64 & 28 & 331 & 34 & \multirow{4}{*}{\bf -} \\
         & speed & 105 & 117 & 112 & 116 &   \\
         &  CPU use & 391 & 387 & 390 & 390 &  \\
         &  power & 9.22 & 8.81 & 9.12 & 9.03 &  \\
         &  battery & $\le 5$ & $\le 2$ & 18 & $\le 2$ &  \\
         &  energy & $\le 80$ & $\le 32$ & 282 & $\le 32$ &  \\
         \midrule
         \multirow{4}{*}{\bf Decode} & len & 423 & 126 & 1012 & 116 &  \textbf{3}-\textbf{7} $\times$\\
         & speed & 37 & 39 & 34 & 39 & \textbf{1/3} \\
         &  CPU use & 393 & 388 & 391  & 391 &  \\
         &  power & 9.43 & 8.97 & 9.68 & 9.14 &  \\
         & battery& 130 & 35 & 301 & 33 & \textbf{16}-\textbf{26} $\times$ \\
         & energy & 1940 & 532 & 4617 & 498 & \textbf{16}-\textbf{26} $\times$ \\
    \bottomrule[1.5pt]
    \end{tabular}
    }
    \vspace{-0.5em}
    \label{tab:decode-dom}
\end{table}



\textbf{Hyper-parameters in power heuristic.}
Table \ref{tab:hyper-param} lists 3 hyper parameters used in our power heuristic. Based on preliminary experiments, EECCS results are not sensitive to their value, which can be explained by Theorem \ref{th:heuristics-acc}.

\begin{table}[!t]
    \centering
    \resizebox{\columnwidth}{!}{
    \begin{tabular}{ccc}
    \toprule[1.5pt]
        \textbf{Hyper Parameter} & \textbf{Definition} & \textbf{Value} \\
    \midrule
       $a_{s}$  & power estimation factor of efficient cores  &  80 \\
       $a_{m}$  & power estimation factor of performance cores & 160 \\
       $a_{b}$  & power estimation factor of prime cores & 200 \\
       $b$  & core idle factor & 0.7\\
       $Ps$   & system static power & 1000\\
    \bottomrule[1.5pt]
    \end{tabular}
    }
    \caption{Hyper-parameters' values in power heuristic.}
    \label{tab:hyper-param}
    \Description{See Caption}
\end{table}


\maketitle

\section{Magnitude of LLM Decoding Current, Power, and Temperature}
In order to supplement the energy consumption magnitude in the main article, we test and list example magnitude of current, power and temperature in Table \ref{tab:energy-challenge2} for reference.

\section{Theoretical Analysis of Power Heuristic Effectiveness}
The following 2 theorems indicates that as long as the magnitude relationship of h(I) is the same as that of power, the accuracy of the final result will be improved, and the variance will be reduced. We denote $energy(I)$ as $e(I)$, and assume observed test profile $e(I)$ is an unbiased observation of ideal energy function $E(I)$.

\begin{theorem}
\label{th:heuristics-acc}

Higher probability of outputting the optimal result.

\noindent If \ $\forall I,J:E(I) > E(J)\iff h(I) > h(J)$,

\noindent then,
\begin{equation*}
\small
    \begin{aligned}
        &P[obj(I)>obj(J)\ |\  E(I)>E(J)]\\
        =&P[e(I)-e(J)>-\frac{1-\alpha}{\alpha}(h(I)-h(J))\ |\  E(I)>E(J)]\\
        >&P[e(I)>e(J)\ |\ E(I)>E(J)]
    \end{aligned}
\end{equation*}

\noindent The weighted averaging objective $obj(I)=(1-\alpha)e(I) + \alpha h(I)t(I)$ leads to higher accuracy than plain $e(I)$ objective.
\end{theorem}

\begin{theorem}
\label{th:heuristics-var}

Variance Reduction.

Since $e(I)$ is measured by $power(I)\cdot t(I)$ in practice, and these 2 terms are independent, we have the following derivation.

\begin{equation*}
\small
    \begin{aligned}
        &Var[(1-\alpha)e(I)+\alpha h(I)t(I)]\\
        =&Var[(1-\alpha)power(I)+\alpha h(I)]\cdot Var[t(I)]\\
        =&(1-\alpha)^2 Var[e(I)] < Var[e(I)]
    \end{aligned}
\end{equation*}
\end{theorem}

\section{Details of Evaluation Platform}

Our evaluation platform: LLM-Profiler features 2 major characteristics, engine extensibility and comprehensive measurement supports.

\textbf{Engine extensibility.} The system currently supports multiple inference engines such as MNN, llama.cpp, mllm, MediaPipe, and executorch. To resolve symbol conflicts such as the one between llama.cpp and mllm, an engine wrapper is abstracted atop all engines, dynamically loading engine libraries during runtime. Any new engines can be integrated by such wrapping interface.

\begin{table}[!t]
    \centering
    \resizebox{0.75\columnwidth}{!}{
    \begin{tabular}{cccc}
    \toprule[1.5pt]
        \multirow{2}{*}{\bf device} & \multicolumn{3}{c}{\bf peak}  \\
        \cmidrule(r){2-4}
        & {\bf current} & {\bf power} & {\bf temperature} \\
        \midrule
        Xiaomi 15 Pro  &  2386 mA & 9.9 W & $80^\circ C$\\
        Mate 40 Pro & 2189 mA & 8.7 W & $85^\circ C$ \\
    \bottomrule[1.5pt]
    \end{tabular}
    }
    \caption{Peak current, power and CPU temperature of Qwen2.5-1.5B (4-bit quantization), running 20 data entries of multi-turn conversation sampled from ShareGPT dataset, with inference engine MNN.}
    \label{tab:energy-challenge2}
    \Description{See Caption}
\end{table}

\begin{figure}[!t]
    \centering
    \includegraphics[width=\linewidth]{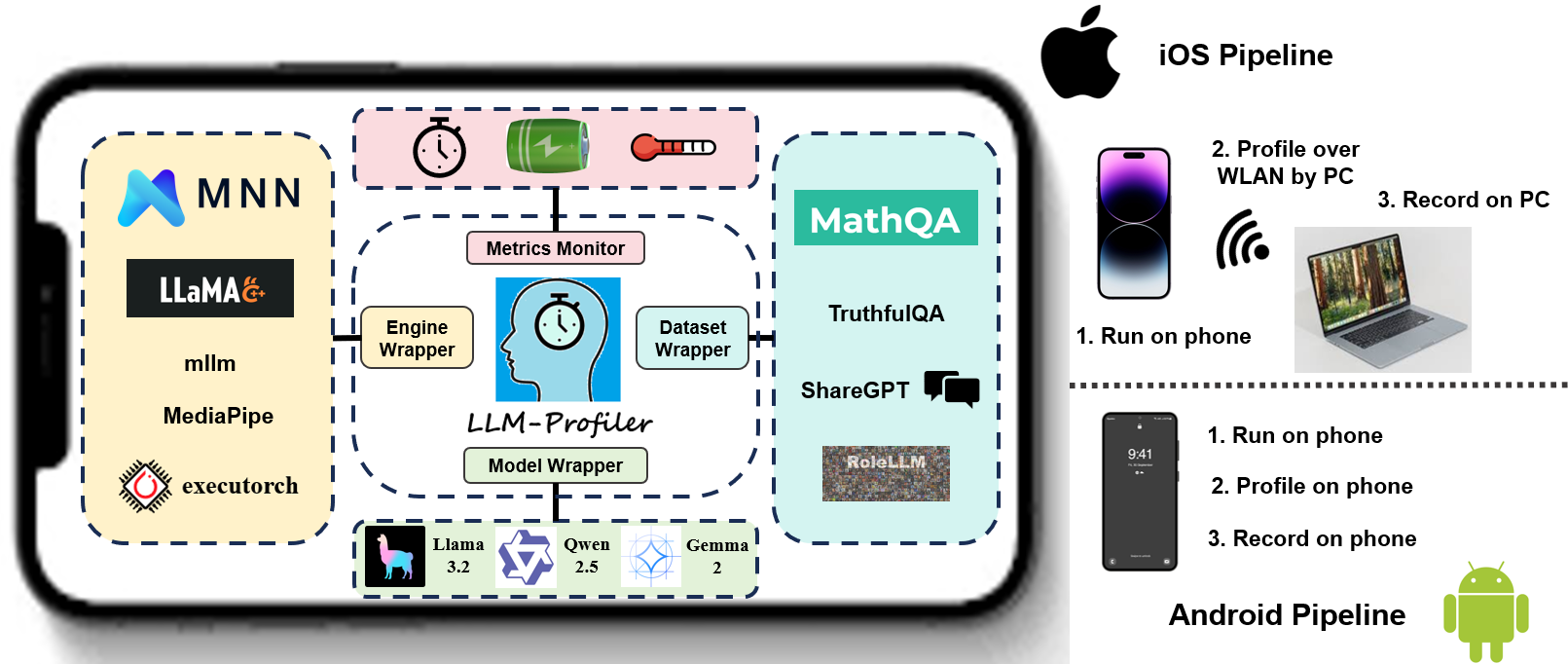}
    \caption{Overview of our evaluation platform \textit{LLM-Profiler}}
    \label{fig:LLM-Profiler}
    \Description{Please see caption.}
\end{figure}

\textbf{Android measurements pipeline.} The whole Android measurements pipeline resides on the phone (Figure \ref{fig:LLM-Profiler}). Speed, energy consumption (current and voltage measurements), battery use (multiplying time and current), and temperature (read from \textit{/sys/class/thermal/thermal\_zone}) are profiled separately for prefill and decode phase. Electricity current data are polled with a time interval of 0.05s, though their values update every 0.25s (1s for Samsung phone) by OS.

\textbf{iOS measurements pipeline.} iOS running and measurements are separated on phone and PC, communicating through tunneld over WLAN (Figure \ref{fig:LLM-Profiler}). Though Sysdiagnose and Xcode energy gauge can both measure energy performance of an iOS app from PC, Sysdiagnose updates energy every 20s and suffers from $\ge10\%$ fluctuation. Therefore, we select Xcode energy gauge \textit{com.apple.xcode.debug-gauge-data-providers.Energy} to get instantaneous power data per second, though such data is only relative power without unit. Besides, CPU utilization rate is also recorded. The pipeline is built upon \textit{pymobiledevice3}.

\section{Experimental Supplementary Details}

\subsection{Engines Versions and Quantization Specification}

\textbf{Engine Versions.} All the engine versions are latest up to February, 2025. We integrate our EECCS into MNN-3.0.4. 3.0.4 is also the version of MNN we compared with. llama.cpp version is b4735, commit hash \textit{73e2ed3c}. mllm version is commit hash \textit{bbf87ffb}. MediaPipe version is 0.10.20. executorch version is 0.5.0, commit hash \textit{e433e610}.

\noindent \textbf{Quantization Specifications.} Because quantization methods impact LLM decoding speed, we carefully choose similar quantization supported across engines to gather fair experimental results.

For MNN, we choose the default quantization: block quantization with block size=128. Qwen-2.5 and Llama-3.2 are quantized into 4 bits, and Gemma-2-2B is quantized into 8 bits to align with MediaPipe.

For llama.cpp, we choose asymmetric 4-bit quantization (Q4\_1) for Qwen-2.5 and Llama-3.2 for fairness, because Q4\_1 has lower computation complexity than MNN block quant. We didn't adopt k-quant because MNN doesn't support super-block, so using it is unfair to llama.cpp.

For mllm, we choose symmetric 4-bit quantization (Q4\_0) for Qwen-2.5 and Llama-3.2 for fairness. mllm only supports Q4\_0 and Q4\_K, so we adopt Q4\_0 for fairness.

For MediaPipe, only asymmetric 8-bit quantized (Q8\_1) Gemma-2-2B model is available.

For executorch, only SpinQuant version Llama-3.2-1B/3B is officially available. We tried Q4\_1 quantization but the models output erroneous texts. Therefore, we use the 4-bit SpinQuant version for comparison. 

\subsection{Statistics of Our Dataset Subsets}
Each subset contains 20 representative samples randomly sampled from the original dataset.

\subsection{Hyper Parameters of EECCS}
\textbf{Slowdown tolerance $\epsilon$.}
We empirically select tolerance $\epsilon=8\%$, which is a slowdown not noticeable to smartphone users, which means the solution is at most 8\% slower than the fastest core selection.

\begin{figure}[!t]
    \centering
    \includegraphics[width=0.95\columnwidth]{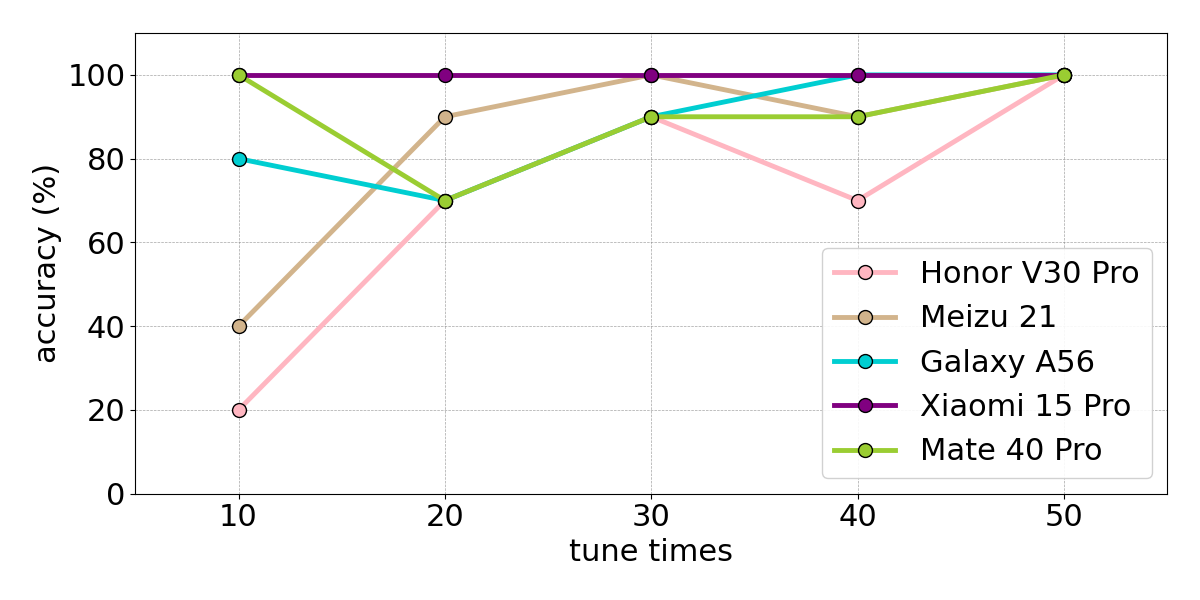}
    \caption{Searching accuracy drops when tune times are small.}
    \label{fig:tuning-times}
    \Description{See Caption}
\end{figure}

\textbf{decode tuning times.}
The default tuning times in EECCS is 50. The experiments in Figure \ref{fig:tuning-times} shows that reducing tuning times to less than 50 leads to lower searching accuracy and suboptimal search results.

\textbf{Hyper-parameters in power heuristic.}
Table \ref{tab:hyper-param} lists 3 hyper parameters used in our power heuristic. Based on preliminary experiments, EECCS results are not sensitive to their value, which can be explained by Theorem \ref{th:heuristics-acc}.

\begin{table}[!t]
    \centering
    \resizebox{\columnwidth}{!}{
\begin{tabular}{ccc}
    \toprule[1.5pt]
        \textbf{Hyper Parameter} & \textbf{Definition} & \textbf{Value} \\
    \midrule
       $a_{s}$  & power estimation factor of efficient cores  &  80 \\
       $a_{m}$  & power estimation factor of performance cores & 160 \\
       $a_{b}$  & power estimation factor of prime cores & 200 \\
       $b$  & core idle factor & 0.7\\
       $Ps$   & system static power & 1000\\
    \bottomrule[1.5pt]
    \end{tabular}
    }
    \caption{Hyper-parameters' values in power heuristic.}
    \label{tab:hyper-param}
    \Description{See Caption}
\end{table}

\subsection{Detailed Results of Fixed Length Tests}

We present all fixed length results in this section, including all 9 combinations of prefill length = 64, 256, 1024, and decode length = 128, 256, 512 on Android and iOS devices.

Our EECCS achieves consistent energy reduction with speed guarantee across all prompt and decode lengths.

\begin{table*}[!t]
    \centering
\resizebox{\linewidth}{!}{
    \begin{tabular}{ccccccccccccccccc}
    \toprule[1.5pt]
    \multirow{2}{*}{\bf device} & \multirow{2}{*}{\bf engine} & \multicolumn{3}{c}{\bf Qwen2.5-1.5B} & \multicolumn{3}{c}{\bf Qwen2.5-3B} & \multicolumn{3}{c}{\bf Llama3.2-1B} & \multicolumn{3}{c}{\bf Llama3.2-3B} & \multicolumn{3}{c}{\bf Gemma2-2B} \\
    \cmidrule(r){3-5}  \cmidrule(r){6-8}  \cmidrule(r){9-11}  \cmidrule(r){12-14} \cmidrule(r){15-17}
         & & {\bf speed} & {\bf battery} & {\bf energy} & {\bf speed} & {\bf battery} & {\bf energy} & {\bf speed} & {\bf battery} & {\bf energy} & {\bf speed} & {\bf battery} & {\bf energy} & {\bf speed} & {\bf battery} & {\bf energy} \\
    \midrule
    \multirow{6}{*}{Mate 40 Pro} & executorch & - & - & - & - & - & - & 14.1 & 34.6 & 505.0 & 6.3 & 88.5 & 1247.5 & - & - & - \\
	 & llama.cpp & 10.2 & 55.0 & 858.2 & 6.7 & 86.5 & 1252.0 & 15.8 & 44.2 & 636.5 & 6.2 & 88.1 & 1331.5 & - & - & - \\
	 & MediaPipe & - & - & - & - & - & - & - & - & - & - & - & - & 4.3 & 121.2 & 1722.2 \\
	 & mllm & 6.1 & 95.0 & 1423.1 & - & - & - & 5.9 & 91.5 & 1421.6 & - & - & - & - & - & - \\
	 & MNN & \textbf{21.7} & 28.9 & 403.1 & \textbf{10.8} & 55.9 & 837.1 & 26.3 & 20.3 & 312.1 & \textbf{10.1} & 59.0 & 884.3 & \textbf{12.7} & 44.1 & 695.1 \\
	 & ours & \underline{20.6} & \textbf{21.0} & \textbf{300.8} & \underline{10.5} & \textbf{40.6} & \textbf{595.5} & \underline{\textbf{26.6}} & \textbf{15.0} & \textbf{227.9} & \underline{9.6} & \textbf{44.5} & \textbf{669.2} & \underline{12.4} & \textbf{33.3} & \textbf{490.0} \\
\end{tabular}
    }
    \caption{Android fixed length test 64-128}
    \label{tab:64-128}
\end{table*}
\begin{table*}[!t]
    \centering
\resizebox{\linewidth}{!}{
    \begin{tabular}{ccccccccccccccccc}
    \toprule[1.5pt]
    \multirow{2}{*}{\bf device} & \multirow{2}{*}{\bf engine} & \multicolumn{3}{c}{\bf Qwen2.5-1.5B} & \multicolumn{3}{c}{\bf Qwen2.5-3B} & \multicolumn{3}{c}{\bf Llama3.2-1B} & \multicolumn{3}{c}{\bf Llama3.2-3B} & \multicolumn{3}{c}{\bf Gemma2-2B} \\
    \cmidrule(r){3-5}  \cmidrule(r){6-8}  \cmidrule(r){9-11}  \cmidrule(r){12-14} \cmidrule(r){15-17}
         & & {\bf speed} & {\bf battery} & {\bf energy} & {\bf speed} & {\bf battery} & {\bf energy} & {\bf speed} & {\bf battery} & {\bf energy} & {\bf speed} & {\bf battery} & {\bf energy} & {\bf speed} & {\bf battery} & {\bf energy} \\
    \midrule
    \multirow{5}{*}{Mate 40 Pro} & executorch & - & - & - & - & - & - & 14.6 & 34.9 & 507.4 & 5.9 & 88.7 & 1246.5 & - & - & - \\
	 & llama.cpp & 10.0 & 52.3 & 816.0 & 6.2 & 84.9 & 1207.7 & 15.5 & 41.3 & 591.0 & 6.0 & 81.8 & 1215.2 & - & - & - \\
	 & mllm & 5.9 & 89.4 & 1316.9 & - & - & - & 5.9 & 86.4 & 1327.4 & - & - & - & - & - & - \\
	 & MNN & \textbf{20.9} & 30.0 & 416.5 & 10.2 & 55.2 & 856.5 & \textbf{26.8} & 20.6 & 317.1 & \textbf{9.8} & 57.4 & 851.8 & \textbf{12.2} & 44.5 & 693.2 \\
	 & ours & \underline{20.6} & \textbf{20.9} & \textbf{297.6} & \underline{\textbf{10.3}} & \textbf{40.7} & \textbf{589.5} & \underline{25.9} & \textbf{15.3} & \textbf{232.1} & \underline{9.5} & \textbf{43.6} & \textbf{644.5} & \underline{12.1} & \textbf{34.2} & \textbf{503.6} \\
	\midrule
	\multirow{5}{*}{V30 Pro} & executorch & - & - & - & - & - & - & 17.6 & 25.4 & 374.9 & 6.8 & 70.6 & 1017.1 & - & - & - \\
\end{tabular}
    }
    \caption{Android fixed length test 64-256}
    \label{tab:64-256}
\end{table*}
\begin{table*}[!t]
    \centering
\resizebox{\linewidth}{!}{
    \begin{tabular}{ccccccccccccccccc}
    \toprule[1.5pt]
    \multirow{2}{*}{\bf device} & \multirow{2}{*}{\bf engine} & \multicolumn{3}{c}{\bf Qwen2.5-1.5B} & \multicolumn{3}{c}{\bf Qwen2.5-3B} & \multicolumn{3}{c}{\bf Llama3.2-1B} & \multicolumn{3}{c}{\bf Llama3.2-3B} & \multicolumn{3}{c}{\bf Gemma2-2B} \\
    \cmidrule(r){3-5}  \cmidrule(r){6-8}  \cmidrule(r){9-11}  \cmidrule(r){12-14} \cmidrule(r){15-17}
         & & {\bf speed} & {\bf battery} & {\bf energy} & {\bf speed} & {\bf battery} & {\bf energy} & {\bf speed} & {\bf battery} & {\bf energy} & {\bf speed} & {\bf battery} & {\bf energy} & {\bf speed} & {\bf battery} & {\bf energy} \\
    \midrule
    \multirow{5}{*}{Mate 40 Pro} & executorch & - & - & - & - & - & - & 13.1 & 37.9 & 550.0 & 5.1 & 93.3 & 1298.1 & - & - & - \\
	 & llama.cpp & 9.2 & 51.3 & 798.3 & 5.8 & 85.9 & 1183.5 & 14.8 & 38.9 & 549.5 & 5.7 & 87.0 & 1294.2 & - & - & - \\
	 & mllm & 5.6 & 86.9 & 1254.6 & - & - & - & 5.6 & 81.7 & 1245.9 & - & - & - & - & - & - \\
	 & MNN & 19.6 & 30.0 & 416.1 & \textbf{9.9} & 52.2 & 802.2 & \textbf{25.7} & 20.7 & 313.8 & \textbf{9.5} & 54.6 & 794.3 & \textbf{11.7} & 44.0 & 686.6 \\
	 & ours & \underline{\textbf{19.8}} & \textbf{21.6} & \textbf{303.9} & \underline{9.8} & \textbf{42.5} & \textbf{609.5} & \underline{25.2} & \textbf{15.7} & \textbf{236.4} & \underline{9.4} & \textbf{43.4} & \textbf{630.8} & \underline{11.5} & \textbf{36.2} & \textbf{531.2} \\
	\midrule
	\multirow{5}{*}{V30 Pro} & executorch & - & - & - & - & - & - & 15.7 & 28.6 & 422.0 & 6.0 & 80.1 & 1128.8 & - & - & - \\
\end{tabular}
    }
    \caption{Android fixed length test 64-512}
    \label{tab:64-512}
\end{table*}
\begin{table*}[!t]
    \centering
\resizebox{\linewidth}{!}{
    \begin{tabular}{ccccccccccccccccc}
    \toprule[1.5pt]
    \multirow{2}{*}{\bf device} & \multirow{2}{*}{\bf engine} & \multicolumn{3}{c}{\bf Qwen2.5-1.5B} & \multicolumn{3}{c}{\bf Qwen2.5-3B} & \multicolumn{3}{c}{\bf Llama3.2-1B} & \multicolumn{3}{c}{\bf Llama3.2-3B} & \multicolumn{3}{c}{\bf Gemma2-2B} \\
    \cmidrule(r){3-5}  \cmidrule(r){6-8}  \cmidrule(r){9-11}  \cmidrule(r){12-14} \cmidrule(r){15-17}
         & & {\bf speed} & {\bf battery} & {\bf energy} & {\bf speed} & {\bf battery} & {\bf energy} & {\bf speed} & {\bf battery} & {\bf energy} & {\bf speed} & {\bf battery} & {\bf energy} & {\bf speed} & {\bf battery} & {\bf energy} \\
    \midrule
    \multirow{6}{*}{Mate 40 Pro} & executorch & - & - & - & - & - & - & 12.5 & 39.9 & 576.3 & 5.2 & 98.6 & 1361.0 & - & - & - \\
	 & llama.cpp & 9.8 & 61.0 & 835.4 & 6.1 & 83.0 & 1137.6 & 15.2 & 40.0 & 564.9 & 5.8 & 83.1 & 1232.9 & - & - & - \\
	 & MediaPipe & - & - & - & - & - & - & - & - & - & - & - & - & 4.4 & 120.3 & 1678.2 \\
	 & mllm & 5.7 & 93.6 & 1354.5 & - & - & - & 5.7 & 91.3 & 1383.5 & - & - & - & - & - & - \\
	 & MNN & \textbf{20.9} & 30.6 & 426.3 & 10.5 & 54.6 & 828.3 & \textbf{26.3} & 20.8 & 320.5 & 9.3 & 62.5 & 912.4 & \textbf{12.1} & 46.3 & 718.3 \\
	 & ours & \underline{20.3} & \textbf{21.5} & \textbf{302.5} & \underline{\textbf{10.6}} & \textbf{40.2} & \textbf{586.9} & \underline{25.6} & \textbf{15.5} & \textbf{233.9} & \underline{\textbf{9.7}} & \textbf{42.8} & \textbf{625.7} & \underline{12.0} & \textbf{34.6} & \textbf{505.5} \\
\end{tabular}
    }
    \caption{Android fixed length test 256-128}
    \label{tab:256-128}
\end{table*}
\begin{table*}[!t]
    \centering
\resizebox{\linewidth}{!}{
    \begin{tabular}{ccccccccccccccccc}
    \toprule[1.5pt]
    \multirow{2}{*}{\bf device} & \multirow{2}{*}{\bf engine} & \multicolumn{3}{c}{\bf Qwen2.5-1.5B} & \multicolumn{3}{c}{\bf Qwen2.5-3B} & \multicolumn{3}{c}{\bf Llama3.2-1B} & \multicolumn{3}{c}{\bf Llama3.2-3B} & \multicolumn{3}{c}{\bf Gemma2-2B} \\
    \cmidrule(r){3-5}  \cmidrule(r){6-8}  \cmidrule(r){9-11}  \cmidrule(r){12-14} \cmidrule(r){15-17}
         & & {\bf speed} & {\bf battery} & {\bf energy} & {\bf speed} & {\bf battery} & {\bf energy} & {\bf speed} & {\bf battery} & {\bf energy} & {\bf speed} & {\bf battery} & {\bf energy} & {\bf speed} & {\bf battery} & {\bf energy} \\
    \midrule
    \multirow{5}{*}{Mate 40 Pro} & executorch & - & - & - & - & - & - & 12.3 & 40.5 & 584.7 & 4.8 & 101.7 & 1402.9 & - & - & - \\
	 & llama.cpp & 9.2 & 51.7 & 800.9 & 6.0 & 83.5 & 1147.1 & 14.1 & 40.2 & 571.3 & 5.7 & 83.9 & 1237.5 & - & - & - \\
	 & mllm & 5.6 & 92.1 & 1323.9 & - & - & - & 5.6 & 86.3 & 1301.0 & - & - & - & - & - & - \\
	 & MNN & \textbf{20.3} & 30.4 & 421.9 & 10.2 & 51.9 & 785.2 & 24.9 & 22.3 & 344.3 & - & - & - & 11.8 & 46.0 & 712.8 \\
	 & ours & \underline{19.8} & \textbf{21.7} & \textbf{305.2} & \underline{\textbf{10.2}} & \textbf{40.5} & \textbf{587.4} & \underline{\textbf{25.3}} & \textbf{15.7} & \textbf{234.3} & \underline{\textbf{9.5}} & \textbf{42.8} & \textbf{625.0} & \underline{\textbf{11.9}} & \textbf{35.0} & \textbf{507.0} \\
	\midrule
	\multirow{5}{*}{V30 Pro} & executorch & - & - & - & - & - & - & 14.4 & 31.2 & 454.0 & 5.5 & 85.8 & 1200.2 & - & - & - \\
\end{tabular}
    }
    \caption{Android fixed length test 256-256}
    \label{tab:256-256}
\end{table*}
\begin{table*}[!t]
    \centering
\resizebox{\linewidth}{!}{
    \begin{tabular}{ccccccccccccccccc}
    \toprule[1.5pt]
    \multirow{2}{*}{\bf device} & \multirow{2}{*}{\bf engine} & \multicolumn{3}{c}{\bf Qwen2.5-1.5B} & \multicolumn{3}{c}{\bf Qwen2.5-3B} & \multicolumn{3}{c}{\bf Llama3.2-1B} & \multicolumn{3}{c}{\bf Llama3.2-3B} & \multicolumn{3}{c}{\bf Gemma2-2B} \\
    \cmidrule(r){3-5}  \cmidrule(r){6-8}  \cmidrule(r){9-11}  \cmidrule(r){12-14} \cmidrule(r){15-17}
         & & {\bf speed} & {\bf battery} & {\bf energy} & {\bf speed} & {\bf battery} & {\bf energy} & {\bf speed} & {\bf battery} & {\bf energy} & {\bf speed} & {\bf battery} & {\bf energy} & {\bf speed} & {\bf battery} & {\bf energy} \\
    \midrule
    \multirow{5}{*}{Mate 40 Pro} & executorch & - & - & - & - & - & - & 11.2 & 43.0 & 618.1 & 4.2 & 112.1 & 1643.8 & - & - & - \\
	 & llama.cpp & 9.1 & 50.3 & 777.8 & 5.6 & 84.2 & 1156.4 & 14.3 & 36.8 & 518.0 & 5.3 & 82.8 & 1204.7 & - & - & - \\
	 & mllm & 5.5 & 86.4 & 1222.9 & - & - & - & 5.5 & 83.5 & 1239.7 & - & - & - & - & - & - \\
	 & MNN & \textbf{19.4} & 29.9 & 415.3 & \textbf{9.9} & 50.4 & 766.1 & \textbf{24.7} & 23.6 & 349.2 & 8.9 & 58.2 & 841.8 & 11.3 & 45.3 & 696.0 \\
	 & ours & \underline{19.1} & \textbf{22.3} & \textbf{313.0} & \underline{9.7} & \textbf{43.1} & \textbf{617.5} & \underline{24.5} & \textbf{16.0} & \textbf{241.4} & \underline{\textbf{9.2}} & \textbf{45.2} & \textbf{655.5} & \underline{\textbf{11.4}} & \textbf{36.6} & \textbf{525.4} \\
	\midrule
	\multirow{5}{*}{V30 Pro} & executorch & - & - & - & - & - & - & 12.9 & 34.3 & 502.5 & 5.1 & 92.9 & 1279.6 & - & - & - \\
\end{tabular}
    }
    \caption{Android fixed length test 256-512}
    \label{tab:256-512}
\end{table*}
\begin{table*}[!t]
    \centering
\resizebox{\linewidth}{!}{
    \begin{tabular}{ccccccccccccccccc}
    \toprule[1.5pt]
    \multirow{2}{*}{\bf device} & \multirow{2}{*}{\bf engine} & \multicolumn{3}{c}{\bf Qwen2.5-1.5B} & \multicolumn{3}{c}{\bf Qwen2.5-3B} & \multicolumn{3}{c}{\bf Llama3.2-1B} & \multicolumn{3}{c}{\bf Llama3.2-3B} & \multicolumn{3}{c}{\bf Gemma2-2B} \\
    \cmidrule(r){3-5}  \cmidrule(r){6-8}  \cmidrule(r){9-11}  \cmidrule(r){12-14} \cmidrule(r){15-17}
         & & {\bf speed} & {\bf battery} & {\bf energy} & {\bf speed} & {\bf battery} & {\bf energy} & {\bf speed} & {\bf battery} & {\bf energy} & {\bf speed} & {\bf battery} & {\bf energy} & {\bf speed} & {\bf battery} & {\bf energy} \\
    \midrule
    \multirow{5}{*}{Mate 40 Pro} & executorch & - & - & - & - & - & - & 9.0 & 54.3 & 779.0 & 3.6 & 135.6 & 1982.2 & - & - & - \\
	 & llama.cpp & 8.8 & 57.3 & 783.1 & 5.8 & 82.4 & 1179.5 & 13.6 & 39.2 & 552.3 & 5.1 & 91.0 & 1298.8 & - & - & - \\
	 & mllm & 4.2 & 111.8 & 1567.0 & - & - & - & 4.3 & 102.7 & 1505.1 & - & - & - & - & - & - \\
	 & MNN & \textbf{18.0} & 32.6 & 450.6 & \textbf{9.5} & 53.8 & 812.5 & \textbf{23.7} & 25.3 & 373.8 & \textbf{8.6} & 60.6 & 870.1 & \textbf{10.7} & 48.8 & 751.4 \\
	 & ours & \underline{17.3} & \textbf{25.0} & \textbf{350.1} & \underline{9.2} & \textbf{45.1} & \textbf{643.4} & \underline{22.4} & \textbf{17.4} & \textbf{260.9} & \underline{8.4} & \textbf{49.4} & \textbf{715.4} & \underline{10.4} & \textbf{44.3} & \textbf{615.3} \\
	\midrule
	\multirow{5}{*}{V30 Pro} & executorch & - & - & - & - & - & - & 9.4 & 45.6 & 664.4 & 3.9 & 123.6 & 1721.7 & - & - & - \\
\end{tabular}
    }
    \caption{Android fixed length test 1024-128}
    \label{tab:1024-128}
\end{table*}
\begin{table*}[!t]
    \centering
\resizebox{\linewidth}{!}{
    \begin{tabular}{ccccccccccccccccc}
    \toprule[1.5pt]
    \multirow{2}{*}{\bf device} & \multirow{2}{*}{\bf engine} & \multicolumn{3}{c}{\bf Qwen2.5-1.5B} & \multicolumn{3}{c}{\bf Qwen2.5-3B} & \multicolumn{3}{c}{\bf Llama3.2-1B} & \multicolumn{3}{c}{\bf Llama3.2-3B} & \multicolumn{3}{c}{\bf Gemma2-2B} \\
    \cmidrule(r){3-5}  \cmidrule(r){6-8}  \cmidrule(r){9-11}  \cmidrule(r){12-14} \cmidrule(r){15-17}
         & & {\bf speed} & {\bf battery} & {\bf energy} & {\bf speed} & {\bf battery} & {\bf energy} & {\bf speed} & {\bf battery} & {\bf energy} & {\bf speed} & {\bf battery} & {\bf energy} & {\bf speed} & {\bf battery} & {\bf energy} \\
    \midrule
    \multirow{5}{*}{Mate 40 Pro} & executorch & - & - & - & - & - & - & 8.4 & 58.4 & 834.0 & 3.4 & 135.1 & 1945.3 & - & - & - \\
	 & llama.cpp & 8.6 & 57.9 & 814.2 & 5.5 & 84.0 & 1192.0 & 13.2 & 38.5 & 536.0 & 4.9 & 86.7 & 1228.8 & - & - & - \\
	 & mllm & 3.9 & 116.2 & 1604.7 & - & - & - & 4.4 & 102.7 & 1489.8 & - & - & - & - & - & - \\
	 & MNN & \textbf{17.1} & 32.1 & 444.0 & \textbf{9.3} & 52.0 & 782.0 & \textbf{22.9} & 24.9 & 370.4 & \textbf{8.4} & 59.8 & 853.1 & \textbf{10.6} & 48.5 & 744.3 \\
	 & ours & \underline{16.8} & \textbf{24.8} & \textbf{344.6} & \underline{9.1} & \textbf{46.2} & \textbf{652.9} & \underline{22.2} & \textbf{17.4} & \textbf{260.3} & \underline{8.4} & \textbf{50.6} & \textbf{724.1} & \underline{10.2} & \textbf{45.7} & \textbf{634.9} \\
	\midrule
	\multirow{5}{*}{V30 Pro} & executorch & - & - & - & - & - & - & 9.1 & 47.1 & 685.7 & 3.8 & 125.8 & 1698.5 & - & - & - \\
\end{tabular}
    }
    \caption{Android fixed length test 1024-256}
    \label{tab:1024-256}
\end{table*}
\begin{table*}[!t]
    \centering
\resizebox{\linewidth}{!}{
    \begin{tabular}{ccccccccccccccccc}
    \toprule[1.5pt]
    \multirow{2}{*}{\bf device} & \multirow{2}{*}{\bf engine} & \multicolumn{3}{c}{\bf Qwen2.5-1.5B} & \multicolumn{3}{c}{\bf Qwen2.5-3B} & \multicolumn{3}{c}{\bf Llama3.2-1B} & \multicolumn{3}{c}{\bf Llama3.2-3B} & \multicolumn{3}{c}{\bf Gemma2-2B} \\
    \cmidrule(r){3-5}  \cmidrule(r){6-8}  \cmidrule(r){9-11}  \cmidrule(r){12-14} \cmidrule(r){15-17}
         & & {\bf speed} & {\bf battery} & {\bf energy} & {\bf speed} & {\bf battery} & {\bf energy} & {\bf speed} & {\bf battery} & {\bf energy} & {\bf speed} & {\bf battery} & {\bf energy} & {\bf speed} & {\bf battery} & {\bf energy} \\
    \midrule
    \multirow{5}{*}{Mate 40 Pro} & executorch & - & - & - & - & - & - & 8.4 & 57.4 & 802.8 & 3.3 & 129.6 & 1840.9 & - & - & - \\
	 & llama.cpp & 8.0 & 57.9 & 792.8 & 5.4 & 82.4 & 1157.5 & 12.8 & 39.0 & 542.8 & 4.3 & 102.5 & 1428.9 & - & - & - \\
	 & mllm & 4.8 & 96.0 & 1316.0 & - & - & - & 4.8 & 88.8 & 1280.6 & - & - & - & - & - & - \\
	 & MNN & \textbf{16.5} & 32.1 & 443.0 & \textbf{9.2} & 51.7 & 775.5 & \textbf{21.7} & 24.2 & 359.6 & 8.2 & 59.7 & 840.6 & \textbf{10.5} & 49.1 & 745.4 \\
	 & ours & \underline{16.4} & \textbf{25.4} & \textbf{348.9} & \underline{8.9} & \textbf{47.1} & \textbf{669.4} & \underline{21.2} & \textbf{17.7} & \textbf{264.2} & \underline{\textbf{8.2}} & \textbf{50.9} & \textbf{730.8} & \underline{10.0} & \textbf{47.2} & \textbf{648.5} \\
	\midrule
	\multirow{5}{*}{V30 Pro} & executorch & - & - & - & - & - & - & 8.9 & 49.3 & 712.7 & 3.7 & 133.0 & 1810.0 & - & - & - \\
\end{tabular}
    }
    \caption{Android fixed length test 1024-512}
    \label{tab:1024-512}
\end{table*}

\begin{table}[!h]
    \centering
\resizebox{0.8\linewidth}{!}{
    \begin{tabular}{cccccc}
    \toprule[1.5pt]
    \multirow{2}{*}{\bf device} & \multirow{2}{*}{\bf engine} & \multicolumn{2}{c}{\bf Qwen2.5-1.5B} & \multicolumn{2}{c}{\bf Llama3.2-1B} \\
    \cmidrule(r){3-4}  \cmidrule(r){5-6}
         & & {\bf speed} & {\bf energy} & {\bf speed} & {\bf energy} \\
    \midrule
    \multirow{3}{*}{iPhone 12} & llama.cpp & 15.3 & 62.3 & 18.9 & 49.3 \\
	 & MNN & 27.6 & 31.5 & 35.1 & 23.7 \\
	 & ours & \underline{\textbf{31.5}} & \textbf{16.1} & \underline{\textbf{40.2}} & \textbf{12.2} \\
	\midrule
	\multirow{3}{*}{iPhone 15} & llama.cpp & 19.5 & 51.3 & 25.6 & 35.3 \\
	 & MNN & 37.2 & 23.4 & 48.5 & 18.0 \\
	 & ours & \underline{\textbf{40.2}} & \textbf{18.6} & \underline{\textbf{52.1}} & \textbf{11.9} \\
    \bottomrule[1.5pt]
    \end{tabular}
    }
    \caption{iOS fixed length test 64-128}
    \label{tab:ios-64-128}
\end{table}
\begin{table}[!t]
    \centering
\resizebox{0.8\linewidth}{!}{
    \begin{tabular}{cccccc}
    \toprule[1.5pt]
    \multirow{2}{*}{\bf device} & \multirow{2}{*}{\bf engine} & \multicolumn{2}{c}{\bf Qwen2.5-1.5B} & \multicolumn{2}{c}{\bf Llama3.2-1B} \\
    \cmidrule(r){3-4}  \cmidrule(r){5-6}
         & & {\bf speed} & {\bf energy} & {\bf speed} & {\bf energy} \\
    \midrule
    \multirow{3}{*}{iPhone 12} & llama.cpp & 14.0 & 70.0 & 18.0 & 54.2 \\
	 & MNN & 27.2 & 32.7 & 34.6 & 28.3 \\
	 & ours & \underline{\textbf{29.8}} & \textbf{17.5} & \underline{\textbf{38.3}} & \textbf{13.3} \\
	\midrule
	\multirow{3}{*}{iPhone 15} & llama.cpp & 18.0 & 54.3 & 23.2 & 41.1 \\
	 & MNN & 35.9 & 24.9 & 46.9 & 19.4 \\
	 & ours & \underline{\textbf{38.2}} & \textbf{19.3} & \underline{\textbf{50.8}} & \textbf{14.6} \\
    \bottomrule[1.5pt]
    \end{tabular}
    }
    \caption{iOS fixed length test 64-256}
    \label{tab:ios-64-256}
\end{table}
\begin{table}[!t]
    \centering
\resizebox{0.8\linewidth}{!}{
    \begin{tabular}{cccccc}
    \toprule[1.5pt]
    \multirow{2}{*}{\bf device} & \multirow{2}{*}{\bf engine} & \multicolumn{2}{c}{\bf Qwen2.5-1.5B} & \multicolumn{2}{c}{\bf Llama3.2-1B} \\
    \cmidrule(r){3-4}  \cmidrule(r){5-6}
         & & {\bf speed} & {\bf energy} & {\bf speed} & {\bf energy} \\
    \midrule
    \multirow{3}{*}{iPhone 12} & llama.cpp & 13.0 & 76.3 & 15.9 & 61.0 \\
	 & MNN & 26.0 & 38.4 & 33.8 & 28.5 \\
	 & ours & \underline{\textbf{27.5}} & \textbf{23.6} & \underline{\textbf{35.4}} & \textbf{20.4} \\
	\midrule
	\multirow{3}{*}{iPhone 15} & llama.cpp & 16.3 & 60.4 & 20.7 & 47.6 \\
	 & MNN & 33.0 & 29.1 & 43.8 & 21.9 \\
	 & ours & \underline{\textbf{34.7}} & \textbf{25.9} & \underline{\textbf{46.4}} & \textbf{18.5} \\
    \bottomrule[1.5pt]
    \end{tabular}
    }
    \caption{iOS fixed length test 64-512}
    \label{tab:ios-64-512}
\end{table}
\begin{table}[!t]
    \centering
\resizebox{0.8\linewidth}{!}{
    \begin{tabular}{cccccc}
    \toprule[1.5pt]
    \multirow{2}{*}{\bf device} & \multirow{2}{*}{\bf engine} & \multicolumn{2}{c}{\bf Qwen2.5-1.5B} & \multicolumn{2}{c}{\bf Llama3.2-1B} \\
    \cmidrule(r){3-4}  \cmidrule(r){5-6}
         & & {\bf speed} & {\bf energy} & {\bf speed} & {\bf energy} \\
    \midrule
    \multirow{3}{*}{iPhone 12} & llama.cpp & 14.3 & 65.2 & 15.9 & 55.2 \\
	 & MNN & 27.0 & 33.5 & 34.5 & 26.4 \\
	 & ours & \underline{\textbf{29.1}} & \textbf{21.0} & \underline{\textbf{38.0}} & \textbf{17.3} \\
	\midrule
	\multirow{3}{*}{iPhone 15} & llama.cpp & 18.7 & 49.9 & 23.6 & 37.7 \\
	 & MNN & 35.5 & 25.3 & 46.7 & 18.7 \\
	 & ours & \underline{\textbf{37.9}} & \textbf{20.0} & \underline{\textbf{49.3}} & \textbf{16.7} \\
    \bottomrule[1.5pt]
    \end{tabular}
    }
    \caption{iOS fixed length test 256-128}
    \label{tab:ios-256-128}
\end{table}
\begin{table}[!t]
    \centering
\resizebox{0.8\linewidth}{!}{
    \begin{tabular}{cccccc}
    \toprule[1.5pt]
    \multirow{2}{*}{\bf device} & \multirow{2}{*}{\bf engine} & \multicolumn{2}{c}{\bf Qwen2.5-1.5B} & \multicolumn{2}{c}{\bf Llama3.2-1B} \\
    \cmidrule(r){3-4}  \cmidrule(r){5-6}
         & & {\bf speed} & {\bf energy} & {\bf speed} & {\bf energy} \\
    \midrule
    \multirow{3}{*}{iPhone 12} & llama.cpp & 13.3 & 72.5 & 17.1 & 56.1 \\
	 & MNN & 26.5 & 36.9 & 34.2 & 28.7 \\
	 & ours & \underline{\textbf{27.8}} & \textbf{20.8} & \underline{\textbf{36.1}} & \textbf{15.8} \\
	\midrule
	\multirow{3}{*}{iPhone 15} & llama.cpp & 17.2 & 56.3 & 22.1 & 42.5 \\
	 & MNN & 34.5 & 28.2 & 45.6 & 21.2 \\
	 & ours & \underline{\textbf{36.1}} & \textbf{22.3} & \underline{\textbf{48.6}} & \textbf{16.1} \\
    \bottomrule[1.5pt]
    \end{tabular}
    }
    \caption{iOS fixed length test 256-256}
    \label{tab:ios-256-256}
\end{table}
\begin{table}[!t]
    \centering
\resizebox{0.8\linewidth}{!}{
    \begin{tabular}{cccccc}
    \toprule[1.5pt]
    \multirow{2}{*}{\bf device} & \multirow{2}{*}{\bf engine} & \multicolumn{2}{c}{\bf Qwen2.5-1.5B} & \multicolumn{2}{c}{\bf Llama3.2-1B} \\
    \cmidrule(r){3-4}  \cmidrule(r){5-6}
         & & {\bf speed} & {\bf energy} & {\bf speed} & {\bf energy} \\
    \midrule
    \multirow{3}{*}{iPhone 12} & llama.cpp & 12.4 & 79.4 & 15.1 & 64.5 \\
	 & MNN & 25.5 & 38.1 & 33.0 & 29.9 \\
	 & ours & \underline{\textbf{26.0}} & \textbf{27.1} & \underline{\textbf{34.1}} & \textbf{21.4} \\
	\midrule
	\multirow{3}{*}{iPhone 15} & llama.cpp & 15.8 & 62.1 & 19.8 & 49.5 \\
	 & MNN & 32.1 & 30.0 & 42.8 & 23.2 \\
	 & ours & \underline{\textbf{33.5}} & \textbf{27.0} & \underline{\textbf{44.7}} & \textbf{19.3} \\
    \bottomrule[1.5pt]
    \end{tabular}
    }
    \caption{iOS fixed length test 256-512}
    \label{tab:ios-256-512}
\end{table}
\begin{table}[!t]
    \centering
\resizebox{0.8\linewidth}{!}{
    \begin{tabular}{cccccc}
    \toprule[1.5pt]
    \multirow{2}{*}{\bf device} & \multirow{2}{*}{\bf engine} & \multicolumn{2}{c}{\bf Qwen2.5-1.5B} & \multicolumn{2}{c}{\bf Llama3.2-1B} \\
    \cmidrule(r){3-4}  \cmidrule(r){5-6}
         & & {\bf speed} & {\bf energy} & {\bf speed} & {\bf energy} \\
    \midrule
    \multirow{3}{*}{iPhone 12} & llama.cpp & 12.0 & 75.8 & 15.1 & 60.5 \\
	 & MNN & \textbf{24.5} & 38.8 & \textbf{31.9} & 30.8 \\
	 & ours & \underline{23.4} & \textbf{34.0} & \underline{30.0} & \textbf{25.5} \\
	\midrule
	\multirow{3}{*}{iPhone 15} & llama.cpp & 16.0 & 58.2 & 20.1 & 44.7 \\
	 & MNN & 31.2 & 30.7 & 41.4 & 22.8 \\
	 & ours & \underline{\textbf{32.4}} & \textbf{29.1} & \underline{\textbf{41.6}} & \textbf{21.1} \\
    \bottomrule[1.5pt]
    \end{tabular}
    }
    \caption{iOS fixed length test 1024-128}
    \label{tab:ios-1024-128}
\end{table}
\begin{table}[!t]
    \centering
\resizebox{0.8\linewidth}{!}{
    \begin{tabular}{cccccc}
    \toprule[1.5pt]
    \multirow{2}{*}{\bf device} & \multirow{2}{*}{\bf engine} & \multicolumn{2}{c}{\bf Qwen2.5-1.5B} & \multicolumn{2}{c}{\bf Llama3.2-1B} \\
    \cmidrule(r){3-4}  \cmidrule(r){5-6}
         & & {\bf speed} & {\bf energy} & {\bf speed} & {\bf energy} \\
    \midrule
    \multirow{3}{*}{iPhone 12} & llama.cpp & 11.6 & 80.9 & 14.8 & 61.4 \\
	 & MNN & \textbf{23.8} & 41.0 & \textbf{31.1} & 32.0 \\
	 & ours & \underline{22.2} & \textbf{33.8} & \underline{29.6} & \textbf{23.5} \\
	\midrule
	\multirow{3}{*}{iPhone 15} & llama.cpp & 15.9 & 59.7 & 19.3 & 48.0 \\
	 & MNN & 29.6 & 32.6 & 39.3 & 24.3 \\
	 & ours & \underline{\textbf{30.4}} & \textbf{29.4} & \underline{\textbf{40.8}} & \textbf{21.2} \\
    \bottomrule[1.5pt]
    \end{tabular}
    }
    \caption{iOS fixed length test 1024-256}
    \label{tab:ios-1024-256}
\end{table}
\begin{table}[!t]
    \centering
\resizebox{0.8\linewidth}{!}{
    \begin{tabular}{cccccc}
    \toprule[1.5pt]
    \multirow{2}{*}{\bf device} & \multirow{2}{*}{\bf engine} & \multicolumn{2}{c}{\bf Qwen2.5-1.5B} & \multicolumn{2}{c}{\bf Llama3.2-1B} \\
    \cmidrule(r){3-4}  \cmidrule(r){5-6}
         & & {\bf speed} & {\bf energy} & {\bf speed} & {\bf energy} \\
    \midrule
    \multirow{3}{*}{iPhone 12} & llama.cpp & 11.0 & 89.1 & 14.0 & 68.9 \\
	 & MNN & \textbf{23.2} & 43.2 & \textbf{30.4} & 33.1 \\
	 & ours & \underline{22.6} & \textbf{34.8} & \underline{29.2} & \textbf{26.9} \\
	\midrule
	\multirow{3}{*}{iPhone 15} & llama.cpp & 14.2 & 68.4 & 17.5 & 55.6 \\
	 & MNN & 27.6 & 35.6 & \textbf{36.7} & 26.6 \\
	 & ours & \underline{\textbf{30.2}} & \textbf{29.6} & \underline{36.3} & \textbf{25.3} \\
    \bottomrule[1.5pt]
    \end{tabular}
    }
    \caption{iOS fixed length test 1024-512}
    \label{tab:ios-1024-512}
\end{table}

\subsection{Detailed Results of Dataset Tests}

We present all dataset test results in this section, including all 4 datasets on Android and iOS devices.

\begin{table*}[!t]
    \centering
\resizebox{\linewidth}{!}{
    \begin{tabular}{ccccccccccccccccc}
    \toprule[1.5pt]
    \multirow{2}{*}{\bf device} & \multirow{2}{*}{\bf engine} & \multicolumn{3}{c}{\bf Qwen2.5-1.5B} & \multicolumn{3}{c}{\bf Qwen2.5-3B} & \multicolumn{3}{c}{\bf Llama3.2-1B} & \multicolumn{3}{c}{\bf Llama3.2-3B} & \multicolumn{3}{c}{\bf Gemma2-2B} \\
    \cmidrule(r){3-5}  \cmidrule(r){6-8}  \cmidrule(r){9-11}  \cmidrule(r){12-14} \cmidrule(r){15-17}
         & & {\bf speed} & {\bf battery} & {\bf energy} & {\bf speed} & {\bf battery} & {\bf energy} & {\bf speed} & {\bf battery} & {\bf energy} & {\bf speed} & {\bf battery} & {\bf energy} & {\bf speed} & {\bf battery} & {\bf energy} \\
    \midrule
    \multirow{6}{*}{Mate 40 Pro} & executorch & - & - & - & - & - & - & 14.0 & 34.6 & 501.4 & 6.0 & 83.9 & 1186.4 & - & - & - \\
	 & llama.cpp & 9.7 & 46.6 & 713.8 & 6.3 & 77.7 & 1112.5 & 15.7 & 35.7 & 516.0 & 6.1 & 83.6 & 1195.3 & - & - & - \\
	 & MediaPipe & - & - & - & - & - & - & - & - & - & - & - & - & 0.5 & 908.9 & 13231.6 \\
	 & mllm & - & - & - & - & - & - & 7.2 & 70.1 & 1086.3 & - & - & - & - & - & - \\
	 & MNN & \textbf{20.5} & 26.9 & 390.5 & 10.5 & 49.8 & 723.0 & \textbf{26.4} & 22.0 & 323.7 & \textbf{9.8} & 56.1 & 787.3 & \textbf{12.5} & 42.7 & 638.8 \\
	 & ours & \underline{19.9} & \textbf{19.7} & \textbf{300.9} & \underline{\textbf{10.6}} & \textbf{41.4} & \textbf{584.3} & \underline{25.7} & \textbf{16.7} & \textbf{231.7} & \underline{9.6} & \textbf{43.5} & \textbf{637.5} & \underline{12.4} & \textbf{34.8} & \textbf{493.7} \\
\end{tabular}
    }
    \caption{Android dataset test MathQA}
    \label{tab:math_qa}
\end{table*}
\begin{table*}[!t]
    \centering
\resizebox{\linewidth}{!}{
    \begin{tabular}{ccccccccccccccccc}
    \toprule[1.5pt]
    \multirow{2}{*}{\bf device} & \multirow{2}{*}{\bf engine} & \multicolumn{3}{c}{\bf Qwen2.5-1.5B} & \multicolumn{3}{c}{\bf Qwen2.5-3B} & \multicolumn{3}{c}{\bf Llama3.2-1B} & \multicolumn{3}{c}{\bf Llama3.2-3B} & \multicolumn{3}{c}{\bf Gemma2-2B} \\
    \cmidrule(r){3-5}  \cmidrule(r){6-8}  \cmidrule(r){9-11}  \cmidrule(r){12-14} \cmidrule(r){15-17}
         & & {\bf speed} & {\bf battery} & {\bf energy} & {\bf speed} & {\bf battery} & {\bf energy} & {\bf speed} & {\bf battery} & {\bf energy} & {\bf speed} & {\bf battery} & {\bf energy} & {\bf speed} & {\bf battery} & {\bf energy} \\
    \midrule
    \multirow{6}{*}{Mate 40 Pro} & executorch & - & - & - & - & - & - & 16.2 & 29.6 & 461.0 & 5.9 & 82.0 & 1192.3 & - & - & - \\
	 & llama.cpp & 10.7 & 54.1 & 806.4 & 6.6 & 80.7 & 1160.5 & 17.0 & 36.3 & 517.0 & 6.0 & 81.9 & 1211.4 & - & - & - \\
	 & MediaPipe & - & - & - & - & - & - & - & - & - & - & - & - & 4.4 & 106.7 & 1543.8 \\
	 & mllm & - & - & - & - & - & - & 7.4 & 72.7 & 1067.9 & - & - & - & - & - & - \\
	 & MNN & \textbf{21.4} & 28.2 & 399.9 & 10.7 & 51.9 & 775.7 & \textbf{28.0} & 20.0 & 292.6 & \textbf{10.1} & 52.1 & 801.6 & 12.0 & 43.7 & 672.6 \\
	 & ours & \underline{20.7} & \textbf{20.0} & \textbf{298.8} & \underline{\textbf{10.8}} & \textbf{35.8} & \textbf{564.1} & \underline{27.7} & \textbf{16.7} & \textbf{233.2} & \underline{9.7} & \textbf{45.7} & \textbf{638.5} & \underline{\textbf{12.0}} & \textbf{37.1} & \textbf{544.6} \\
\end{tabular}
    }
    \caption{Android dataset test RoleBench}
    \label{tab:RoleBench}
\end{table*}
\begin{table*}[!t]
    \centering
\resizebox{\linewidth}{!}{
    \begin{tabular}{ccccccccccccccccc}
    \toprule[1.5pt]
    \multirow{2}{*}{\bf device} & \multirow{2}{*}{\bf engine} & \multicolumn{3}{c}{\bf Qwen2.5-1.5B} & \multicolumn{3}{c}{\bf Qwen2.5-3B} & \multicolumn{3}{c}{\bf Llama3.2-1B} & \multicolumn{3}{c}{\bf Llama3.2-3B} & \multicolumn{3}{c}{\bf Gemma2-2B} \\
    \cmidrule(r){3-5}  \cmidrule(r){6-8}  \cmidrule(r){9-11}  \cmidrule(r){12-14} \cmidrule(r){15-17}
         & & {\bf speed} & {\bf battery} & {\bf energy} & {\bf speed} & {\bf battery} & {\bf energy} & {\bf speed} & {\bf battery} & {\bf energy} & {\bf speed} & {\bf battery} & {\bf energy} & {\bf speed} & {\bf battery} & {\bf energy} \\
    \midrule
    \multirow{6}{*}{Mate 40 Pro} & executorch & - & - & - & - & - & - & 17.2 & 28.2 & 427.9 & 6.6 & 80.6 & 1121.6 & - & - & - \\
	 & llama.cpp & 10.9 & 59.0 & 811.6 & 6.9 & 74.0 & 1147.1 & 16.9 & 35.2 & 496.8 & 6.0 & 86.3 & 1230.3 & - & - & - \\
	 & MediaPipe & - & - & - & - & - & - & - & - & - & - & - & - & 4.5 & 101.9 & 1538.5 \\
	 & mllm & - & - & - & - & - & - & 7.3 & 76.2 & 1086.3 & - & - & - & - & - & - \\
	 & MNN & \textbf{21.9} & 25.3 & 381.5 & 10.9 & 51.7 & 758.7 & \textbf{27.4} & 21.7 & 313.6 & \textbf{10.2} & 54.3 & 815.5 & \textbf{12.4} & 45.8 & 675.3 \\
	 & ours & \underline{21.8} & \textbf{19.2} & \textbf{280.8} & \underline{\textbf{11.0}} & \textbf{36.6} & \textbf{559.5} & \underline{26.9} & \textbf{14.5} & \textbf{218.2} & \underline{9.8} & \textbf{44.9} & \textbf{633.6} & \underline{12.3} & \textbf{37.2} & \textbf{531.0} \\
\end{tabular}
    }
    \caption{Android dataset test TruthfulQA}
    \label{tab:truthful_qa}
\end{table*}
\begin{table*}[!t]
    \centering
\resizebox{\linewidth}{!}{
    \begin{tabular}{ccccccccccccccccc}
    \toprule[1.5pt]
    \multirow{2}{*}{\bf device} & \multirow{2}{*}{\bf engine} & \multicolumn{3}{c}{\bf Qwen2.5-1.5B} & \multicolumn{3}{c}{\bf Qwen2.5-3B} & \multicolumn{3}{c}{\bf Llama3.2-1B} & \multicolumn{3}{c}{\bf Llama3.2-3B} & \multicolumn{3}{c}{\bf Gemma2-2B} \\
    \cmidrule(r){3-5}  \cmidrule(r){6-8}  \cmidrule(r){9-11}  \cmidrule(r){12-14} \cmidrule(r){15-17}
         & & {\bf speed} & {\bf battery} & {\bf energy} & {\bf speed} & {\bf battery} & {\bf energy} & {\bf speed} & {\bf battery} & {\bf energy} & {\bf speed} & {\bf battery} & {\bf energy} & {\bf speed} & {\bf battery} & {\bf energy} \\
    \midrule
    \multirow{4}{*}{Mate 40 Pro} & executorch & - & - & - & - & - & - & 10.2 & 43.5 & 632.0 & 4.1 & 103.9 & 1486.1 & - & - & - \\
	 & llama.cpp & 9.0 & 52.5 & 756.7 & 5.3 & 78.4 & 1152.1 & 13.6 & 35.7 & 508.7 & 5.0 & 88.1 & 1257.8 & - & - & - \\
	 & MNN & \textbf{19.0} & 26.8 & 385.7 & \textbf{9.8} & 46.5 & 693.3 & \textbf{23.1} & 21.3 & 328.6 & 8.6 & 54.1 & 774.9 & - & - & - \\
	 & ours & \underline{18.9} & \textbf{21.7} & \textbf{312.2} & \underline{9.7} & \textbf{43.5} & \textbf{626.4} & \underline{22.9} & \textbf{17.9} & \textbf{263.9} & \underline{\textbf{8.7}} & \textbf{48.1} & \textbf{691.7} & - & - & - \\
	\midrule
	\multirow{5}{*}{V30 Pro} & executorch & - & - & - & - & - & - & 11.6 & 40.3 & 569.6 & - & - & - & - & - & - \\
	 & llama.cpp & 9.0 & 50.3 & 721.0 & 5.7 & 78.4 & 1133.9 & 14.1 & 35.5 & 485.3 & - & - & - & - & - & - \\
\end{tabular}
    }
    \caption{Android dataset test ShareGPT}
    \label{tab:shareGPT}
\end{table*}

\begin{table}[!t]
    \centering
\resizebox{0.8\linewidth}{!}{
    \begin{tabular}{cccccc}
    \toprule[1.5pt]
    \multirow{2}{*}{\bf device} & \multirow{2}{*}{\bf engine} & \multicolumn{2}{c}{\bf Qwen2.5-1.5B} & \multicolumn{2}{c}{\bf Llama3.2-1B} \\
    \cmidrule(r){3-4}  \cmidrule(r){5-6}
         & & {\bf speed} & {\bf energy} & {\bf speed} & {\bf energy} \\
    \midrule
    \multirow{3}{*}{iPhone 12} & llama.cpp & 12.8 & 76.6 & 16.6 & 58.1 \\
	 & MNN & 28.0 & 34.9 & 34.8 & 27.2 \\
	 & ours & \underline{\textbf{28.9}} & \textbf{23.5} & \underline{\textbf{37.8}} & \textbf{15.5} \\
	\midrule
	\multirow{3}{*}{iPhone 15} & llama.cpp & 16.0 & 61.0 & 20.7 & 46.9 \\
	 & MNN & 34.2 & 28.2 & 46.8 & 20.1 \\
	 & ours & \underline{\textbf{35.3}} & \textbf{24.4} & \underline{\textbf{48.8}} & \textbf{16.5} \\
    \bottomrule[1.5pt]
    \end{tabular}
    }
    \caption{iOS dataset test MathQA}
    \label{tab:ios-math_qa}
\end{table}
\begin{table}[!t]
    \centering
\resizebox{0.8\linewidth}{!}{
    \begin{tabular}{cccccc}
    \toprule[1.5pt]
    \multirow{2}{*}{\bf device} & \multirow{2}{*}{\bf engine} & \multicolumn{2}{c}{\bf Qwen2.5-1.5B} & \multicolumn{2}{c}{\bf Llama3.2-1B} \\
    \cmidrule(r){3-4}  \cmidrule(r){5-6}
         & & {\bf speed} & {\bf energy} & {\bf speed} & {\bf energy} \\
    \midrule
    \multirow{3}{*}{iPhone 12} & llama.cpp & 15.1 & 63.0 & 18.0 & 53.7 \\
	 & MNN & 28.4 & 32.2 & 35.9 & 23.9 \\
	 & ours & \underline{\textbf{31.6}} & \textbf{15.2} & \underline{\textbf{41.1}} & \textbf{12.7} \\
	\midrule
	\multirow{3}{*}{iPhone 15} & llama.cpp & 20.1 & 46.2 & 23.3 & 40.7 \\
	 & MNN & 37.1 & 23.0 & 49.9 & 16.2 \\
	 & ours & \underline{\textbf{40.3}} & \textbf{17.2} & \underline{\textbf{53.2}} & \textbf{12.7} \\
    \bottomrule[1.5pt]
    \end{tabular}
    }
    \caption{iOS dataset test RoleBench}
    \label{tab:ios-RoleBench}
\end{table}
\begin{table}[!t]
    \centering
\resizebox{0.8\linewidth}{!}{
    \begin{tabular}{cccccc}
    \toprule[1.5pt]
    \multirow{2}{*}{\bf device} & \multirow{2}{*}{\bf engine} & \multicolumn{2}{c}{\bf Qwen2.5-1.5B} & \multicolumn{2}{c}{\bf Llama3.2-1B} \\
    \cmidrule(r){3-4}  \cmidrule(r){5-6}
         & & {\bf speed} & {\bf energy} & {\bf speed} & {\bf energy} \\
    \midrule
    \multirow{3}{*}{iPhone 12} & llama.cpp & 15.7 & 59.7 & 17.7 & 53.9 \\
	 & MNN & 28.8 & 31.0 & 35.4 & 26.0 \\
	 & ours & \underline{\textbf{32.0}} & \textbf{14.5} & \underline{\textbf{39.4}} & \textbf{12.8} \\
	\midrule
	\multirow{3}{*}{iPhone 15} & llama.cpp & 20.1 & 45.4 & 23.1 & 40.8 \\
	 & MNN & 37.8 & 22.6 & 48.9 & 17.9 \\
	 & ours & \underline{\textbf{41.0}} & \textbf{16.6} & \underline{\textbf{51.5}} & \textbf{13.6} \\
    \bottomrule[1.5pt]
    \end{tabular}
    }
    \caption{iOS dataset test TruthfulQA}
    \label{tab:ios-truthful_qa}
\end{table}
\begin{table}[!t]
    \centering
\resizebox{0.8\linewidth}{!}{
    \begin{tabular}{cccccc}
    \toprule[1.5pt]
    \multirow{2}{*}{\bf device} & \multirow{2}{*}{\bf engine} & \multicolumn{2}{c}{\bf Qwen2.5-1.5B} & \multicolumn{2}{c}{\bf Llama3.2-1B} \\
    \cmidrule(r){3-4}  \cmidrule(r){5-6}
         & & {\bf speed} & {\bf energy} & {\bf speed} & {\bf energy} \\
    \midrule
    \multirow{3}{*}{iPhone 12} & llama.cpp & 10.2 & 98.0 & 11.8 & 83.8 \\
	 & MNN & \textbf{26.2} & 37.9 & \textbf{31.8} & 31.1 \\
	 & ours & \underline{24.7} & \textbf{25.3} & \underline{31.0} & \textbf{21.1} \\
	\midrule
	\multirow{3}{*}{iPhone 15} & llama.cpp & 14.6 & 68.0 & 14.8 & 67.0 \\
	 & MNN & 24.4 & 40.7 & 38.9 & 25.5 \\
	 & ours & \underline{\textbf{25.3}} & \textbf{35.2} & \underline{\textbf{39.0}} & \textbf{22.3} \\
    \bottomrule[1.5pt]
    \end{tabular}
    }
    \caption{iOS dataset test ShareGPT}
    \label{tab:ios-shareGPT}
\end{table}